\documentclass[journal,12pt,onecolumn,draftclsnofoot,]{IEEEtran}
\IEEEoverridecommandlockouts
\usepackage{cite}
\usepackage{amsmath,amssymb,amsfonts}
\usepackage{algorithmic}
\usepackage{bbm}
\usepackage{graphicx}
\usepackage{diagbox}
\usepackage{textcomp}
\DeclareMathOperator*{\argmax}{argmax} 
\usepackage{xcolor}
\usepackage{scalerel}
\usepackage{lettrine}
\usepackage{dsfont}
\usepackage{subcaption}
\usepackage{float}
\usepackage{amsmath,amsthm}
\usepackage[]{graphicx}
\usepackage{lipsum}
\usepackage[nolist]{acronym}

\def\BibTeX{{\rm B\kern-.05em{\sc i\kern-.025em b}\kern-.08em
    T\kern-.1667em\lower.7ex\hbox{E}\kern-.125emX}}

\begin{document}

\newcommand{\nRx}{\ensuremath{ L}}
\newcommand{\tru}{\ensuremath{ S}}
\newcommand{\truB}{\ensuremath{\tilde{\tru}}}
\newcommand{\maxTru}{\ensuremath{{\tru^*}}}
\newcommand{\truSA}{\ensuremath{ S_{{sa}}}}
\newcommand{\truU}{\ensuremath{ S_{ u}}}
\newcommand{\truD}{\ensuremath{ S_{ d}}}
\newcommand{\truUC}{\ensuremath{ S_{{u,c}}}}
\newcommand{\truUNC}{\ensuremath{ S_{{u,\bar c}}}}
\newcommand{\truDC}{\ensuremath{ S_{{d,c}}}}
\newcommand{\truDNC}{\ensuremath{ S_{{d,\bar c}}}}

\newcommand{\load}{\ensuremath{{G}_{c}}}
\newcommand{\loadNC}{\ensuremath{{G}_{\bar c}}}
\newcommand{\peras}{\ensuremath{\epsilon}}
\newcommand{\perasU}{\ensuremath{\peras_{ 1}}}
\newcommand{\perasD}{\ensuremath{\peras_{ 2}}}
\newcommand{\NTx}{\ensuremath{{N}_{c}}}
\newcommand{\nTx}{{n_{c}}}
\newcommand{\pforw}{\ensuremath{\delta_{\nTx}}}
\newcommand{\NTxNC}{\ensuremath{{N}_{\bar{c}}}}
\newcommand{\nTxNC}{{n_{\bar{c}}}}
\newcommand{\pforwNC}{\ensuremath{\delta_{\nTxNC}}}
\newcommand{\psU}{{p}_{\nTx}}
\newcommand{\psUNC}{{p}_{\nTxNC}}
\newcommand{\psD}{{q}_{\nTx}}
\newcommand{\psDNC}{{q}_{\nTxNC}}
\newcommand{\ancF}{\ensuremath{\mathcal H}} 

\newcommand{\psDP}{\psD'}
\newcommand{\bNC}{\gamma}
\newcommand{\ancA}{\ensuremath{\mathcal A}} 

\begin{acronym}
\acro{ACRDA}{asynchronous contention resolution diversity ALOHA}%
\acro{AP}{access point}%
\acro{AWGN}{additive white Gaussian noise}%
\acro{BS}{base station}%
\acro{CDF}{cumulative distribution function}
\acro{CRA}{contention resolution ALOHA}
\acro{CRDSA}{contention resolution diversity slotted ALOHA}
\acro{CSA}{coded slotted ALOHA}
\acro{CS}{critical service}
\acro{DAMA}{demand assigned multiple access}%
\acro{DSA}{diversity slotted ALOHA}%
\acro{ECRA}{enhanced contention resolution ALOHA}%
\acro{FEC}{forward error correction}%
\acro{GEO}{geostationary orbit}%

\acro{IC}{interference cancellation}%
\acro{IoT}{Internet of Things}%
\acro{IRCRA}{irregular repetition contention resolution ALOHA}%
\acro{IRSA}{irregular repetition slotted ALOHA} %

\acro{LT}{Luby transform}%
\acro{LEO}{Low-Earth Orbit}
\acro{M2M}{machine-to-machine}%
\acro{MAC}{medium access}%
\acro{MF-TDMA}{multi-frequency time division multiple access}%
\acro{MRC}{maximal-ratio combining}
\acro{NOMA}{non-orthogonal multiple access}
\acro{NCS}{non-critical service}
\acro{NTN}{non-terrestrial networks}

\acro{PDF}{probability density function}%
\acro{PER}{packet error rate}%
\acro{PLR}{packet loss rate}%
\acro{PMF}{probability mass function}%

\acro{RA}{random access}%
\acro{SA}{slotted ALOHA}%
\acro{SB}{Shannon bound}%
\acro{SC}{selection combining}%
\acro{SIC}{successive interference cancellation}%
\acro{SINR}{signal-to-interference and noise ratio}%
\acro{SNIR}{signal-to-noise-plus-interference ratio}%
\acro{SNR}{signal-to-noise ratio}%
\acro{TDMA}{time division multiple access}%

\acro{VF}{virtual frame}


\end{acronym}

\title{Grant-Free Coexistence of Critical and Non-Critical IoT Services in Two-Hop Satellite and Terrestrial Networks\\
\thanks{The work of Rahif Kassab and Osvaldo Simeone has received funding from the European  Research  Council (ERC) under the European Union Horizon 2020 research and innovation program (grant agreement 725731).}
}

\author{\IEEEauthorblockN{Rahif Kassab\IEEEauthorrefmark{1}, Andrea Munari\IEEEauthorrefmark{2}, Federico Clazzer\IEEEauthorrefmark{2}, and Osvaldo Simeone\IEEEauthorrefmark{1} }\\
\IEEEauthorblockA{\small\IEEEauthorrefmark{1}King's Communications, Learning and Information Processing (KCLIP) Lab, King's College London, UK\\
\small\IEEEauthorrefmark{2}Institute of Communications and Navigation, German Aerospace Center (DLR),  82234 Wessling, Germany\\
Emails: \IEEEauthorrefmark{1}\{rahif.kassab,osvaldo.simeone\}@kcl.ac.uk , \IEEEauthorrefmark{2}\{Andrea.Munari,Federico.Clazzer\}@dlr.de}}

\maketitle

\begin{abstract}
Terrestrial and satellite communication networks often rely on two-hop wireless architectures with an access channel followed by backhaul links. Examples include Cloud-Radio Access Networks (C-RAN) and Low-Earth Orbit (LEO) satellite systems. Furthermore, communication services characterized by the coexistence of heterogeneous requirements are emerging as key use cases. This paper studies the performance of critical service (CS) and non-critical service (NCS) for Internet of Things (IoT) systems sharing a grant-free channel consisting of radio access and backhaul segments. On the radio access segment, IoT devices send packets to a set of non-cooperative access points (APs) using slotted ALOHA (SA). The APs then forward correctly received messages to a base station over a shared wireless backhaul segment adopting SA. 
We study first a simplified erasure channel model, which is well suited for satellite applications. Then, in order to account for terrestrial scenarios, the impact of fading is considered. 
Among the main conclusions, we show that orthogonal inter-service resource allocation is generally preferred for NCS devices, while non-orthogonal protocols can improve the throughput and packet success rate of CS devices for both terrestrial and satellite scenarios.
\end{abstract}
\begin{IEEEkeywords}
Beyond 5G, IoT, Grant-Free, satellite networks, mMTC, URLLC
\end{IEEEkeywords}
\section{Introduction}
Future generations of cellular and satellite networks 
will include new services with vastly different performance requirements. In recent 3GPP releases \cite{3gpp}, a distinction is made among Ultra-Reliable and Low-Latency Communications (URLLC), with stringent delays and packet success rate requirements; enhanced Mobile Broadband (eMBB) for high throughput; and massive Machine Type Communications (mMTC) for sporadic transmissions with large spatial densities of devices \cite{5Goverview,opportunistic_coexistence}. In this paper, we focus on Internet of Things (IoT) scenarios, which are typically assumed to fall into the mMTC service category \cite{nbiot_mmtc}. We take a further step compared to the mentioned 3GPP classification by considering a beyond-5G scenario characterized by the coexistence of heterogeneous IoT devices having critical and non-critical service requirements. Devices with \ac{CS} requirements must be provided more stringent throughput and packet success rate performance guarantees than \ac{NCS} devices. We note that \ac{CS} for IoT will be introduced in 3GPP release 17 \cite{3gpp_rel_17} under the name \textit{enhanced Industrial IoT}, while \ac{NCS} IoT is a typical use case for \textit{New Radio-light}, which will also be studied in release 17 \cite{3gpp_rel_17}. \par
\begin{figure*}
\centering
\begin{subfigure}{.5\textwidth}
  \centering
	\includegraphics[height= 6.5 cm, width= 7.8 cm]{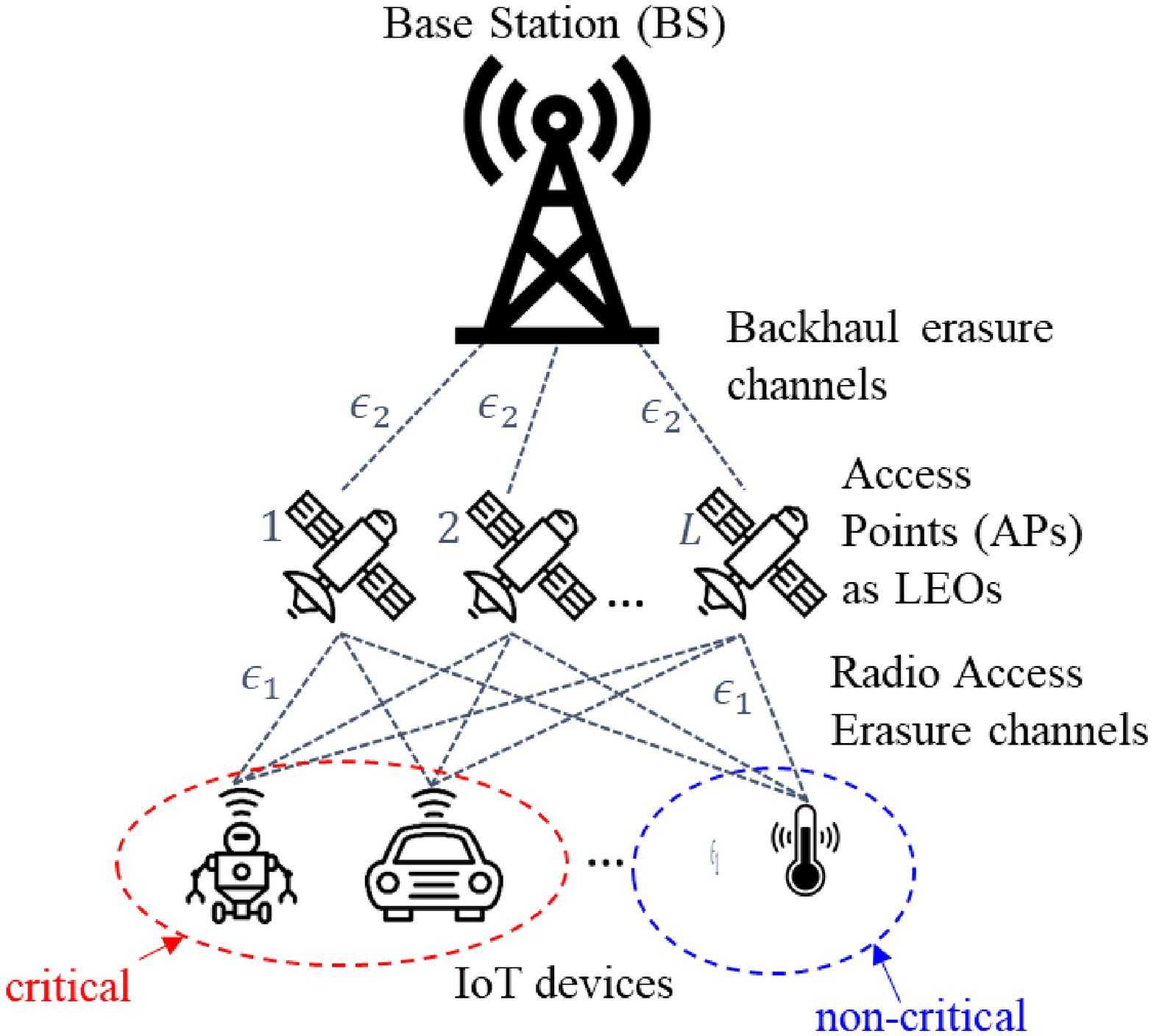}
  \caption{}
  \label{fig:system_model_space}
\end{subfigure}%
\begin{subfigure}{.5\textwidth}
  \centering
	\includegraphics[height= 6.5 cm, width= 7.3 cm]{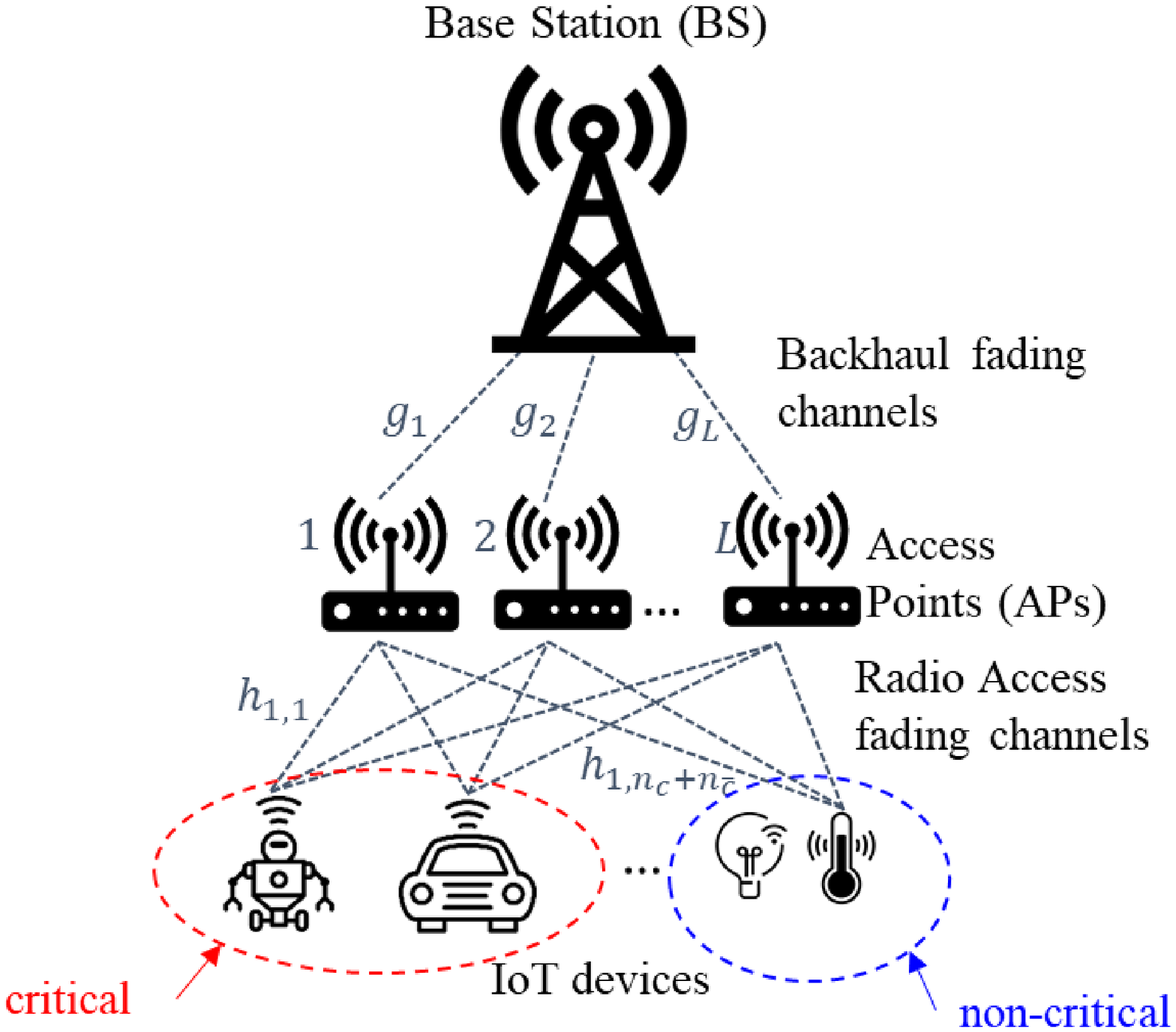}
  \caption{}
  \label{fig:system_model_earth}
\end{subfigure}
\caption{An IoT system with grant-free wireless radio access and shared backhaul with uncoordinated APs, in which IoT devices generate \ac{CS} or \ac{NCS} messages. The setup in (a) illustrates a satellite communications scenario with binary erasure channels modeling the presence/absence of a line-of-sight link. The setup in (b) illustrates a terrestrial communications scenario with fading channels.}
\label{fig:system_model}
\end{figure*}

As illustrated in Fig.~\ref{fig:system_model}, we consider a two-hop wireless architecture with radio access channels followed by backhaul links. In these topologies, space diversity is provided by multiple Access Points (APs) that play the role of relays between the devices and the Base Station (BS). For terrestrial networks, C-RAN can be described as two-hop networks \cite{bookcransimeone} while for satellite networks, the model applies to communications through \ac{LEO} mega-constellations, e.g., Amazon Kuiper \cite{amazon_kuiper} and SpaceX Starlink \cite{spacex_starlink} projects. A new 3GPP work item for IoT over \ac{NTN} was recently introduced \cite{3gpp_rel_17}. With thousands of LEO satellites, these constellations will offer connectivity to each earth location with multiple satellites at a time. Satellite based IoT deployments can hence extend connectivity to remote areas with low or no cellular coverage, enabling applications in many industries, such as maritime and road transportation, farming and mining.
\par
In the presence of a large number of IoT devices requiring the transmission of small amounts of data, conventional \textit{grant-based} radio access protocols can cause a significant overhead on the access network due to the large number of handshakes required. A potentially more efficient solution is given by \textit{grant-free} radio access protocols where devices transmit whenever they have a packet to deliver without any prior handshake \cite{grant_free_popovski,rahif_grant_free,grant_free_cavdar}. This is typically done via some variations of the classical ALOHA random access scheme \cite{abramson1970aloha}. Grant-free access protocols are used by many commercial solutions in the terrestrial domain, e.g., by Sigfox \cite{sigfox} and LoRaWAN \cite{lora}; as well as in the satellite domain, using constellations of low-earth orbit satellites (LEO), e.g. Orbcomm \cite{orbcomm} and Myriota \cite{myriota}.  \par
In classical cellular IoT scenarios, \textit{orthogonal inter-service resource allocation} schemes are typically used \cite{3gpp_nbiot}. However, due to their static nature, orthogonal schemes may cause an inefficient use of resources under unpredictable traffic patterns in grant-free IoT systems. To obviate this problem, 
\textit{non-orthogonal resource allocation}, which allows access of multiple devices to the same time-frequency resource, presents a promising alternative solution \cite{noma_saito,noma_performance,noma_challenges_potential,ding2014performance}. Recent work has proposed to apply non-orthogonal resource allocation to heterogeneous services \cite{rahif_access_2018,popovski2018slicing, rahifuplink}. In order to mitigate the impact of interference in non-orthogonal schemes, one can leverage successive interference cancellation (SIC) \cite{aloha_noma}, time diversity \cite{coded_slotted_aloha}, and/or space diversity \cite{munari_multiple_aloha}\cite{vladimir_cooperative_ALOHA}. \par 
\textit{Main Contributions: }In this work, we aim to answer the following questions: \textit{What is the impact of two-hop topologies on the grant-free coexistence of \ac{CS} and \ac{NCS} in IoT systems? Is there any advantage in using non-orthogonal access techniques across the two services?} In order to do this, we study grant-free access for \ac{CS} and \ac{NCS} in space diversity-based models for both satellite and terrestrial applications. We analytically derive throughput and packet success rate measures for both \ac{CS} and \ac{NCS} as a function of key parameters such as the number of APs and frame size. The analysis, which generalizes conventional models used for \ac{SA}, accounts for orthogonal and non-orthogonal inter-service access schemes, as well as for binary erasure channels modeling \ac{NTN} (Fig.~\ref{fig:system_model}(a)) and for fading channels modeling terrestrial networks (Fig.~\ref{fig:system_model}(b)). Finally, two receiver models are considered, namely, a collision model, where packets transmitted by the same device are assumed to undergo destructive collision, and a superposition model, where packets transmitted from the same device are superposed at the receiver. \par

Preliminary results for our model were presented in \cite{frederico2019modern} and \cite{rahif_space}. \textcolor{black}{The main differences with our two previous works are as follows:
\begin{itemize}
        \item Reference \cite{frederico2019modern} considers a single-service set-up with the erasure channel model and a single and simpler decoder model. In contrast, this work considers the coexistence of two services with both erasure and fading channel models in addition to a more practical decoder model.
        \item Paper \cite{rahif_space} studies CS and NCS model under erasure channels assuming that the decoder cannot recover a packet if multiple instances of the same packet are received. In contrast, this work considers fading channel model in addition to a more practical decoder model.
\end{itemize} 
}
\par 
The rest of the paper is organized as follows. In Sec.~\ref{sec:system_model_performance_metrics} we describe the system model used and the performance metrics. In Sec.~\ref{sec:oneclass} we study the system in the presence of a single service while Sec. \ref{sec:heterogeneous_collision} and \ref{sec:heterogeneous_superposition} tackle the heterogeneous services case under the general collision and superposition model respectively. Finally, the heterogeneous service case is evaluated under the fading channel model in Sec.~\ref{sec:throughput_reliability_fading}, conclusions and extensions are discussed in Sec.~\ref{sec:conclusions}.\par
\textbf{Notation: }Throughout our discussion, we denote as $X\sim \operatorname{Bin}(n,p)$ a Binomial random variable (RV) with $n$ trials and probability of success $p$; as $X ~\sim \operatorname{Poiss}(\lambda)$ a Poisson RV with parameter $\lambda$. We also write $(X,Y)\sim f \cdot g$ for two independent RVs $X$ and $Y$ with respective probability density functions $f$ and $g$.
\section{\textcolor{black}{Related Works}}
\label{sec:related_works}
\textcolor{black}{
\textit{Satellite networks}, through constellations of LEO satellites, have been increasingly receiving interest from academia and industry as a potential technology to extend connectivity to rural and remote areas and thus enabling new applications in fields like maritime, transportation, farming and mining where internet connectivity is a big challenge. A new 3GPP work item for IoT over non-terrestrial networks (NTN) was recently introduced \cite{3gpp_rel_17}. Academic research in the wireless community has been focusing on the integration of satellite networks in next generation cellular networks. For example, \cite{dra_leo} considers the problem dynamic resource allocations for LEO satellites while taking into consideration the power consumption and mobility management constraints. \cite{traffic_leo} analyzes IoT traffic in LEO satellites networks while \cite{ultra_dense_leo, broadband_leo, noma_saito} consider different architectures integrated with 5G and beyond cellular networks while highlighting the advantages of such architectures which include space and access diversity. Similarly to previous works, this work considers a two-hop architecture to model a satellite system, however, in contrast to previous works, we derive the throughput expressions with coexisting critical and non-critical IoT services.}\par
\textcolor{black}{\textit{Non-orthogonal multiple access} is a promising technology to provide massive connectivity by allowing for multiple users to transmit over the same radio resources in the uplink while using superposition coding in the downlink.  For the uplink, reference \cite{ul_noma_2} shows that NOMA with Successive Interference Cancellation (SIC) at the base stations can significantly enhance cell-edge users’ throughput. As for the downlink, NOMA was demonstrated to achieve superior performance in terms of ergodic sum rate of a cellular network with randomly deployed users in \cite{ding2014performance}. NOMA can leverage different technologies to differentiate the users, e.g., in the code or power domains. We refer to \cite{survey_noma_dai} and references therein for a detailed discussion.} \par
\textcolor{black}{Although NOMA can provide massive connectivity, a related challenge is the access to radio resources. Grant-based access, which is currently used in LTE/LTE-A \cite{RA_lte_noma}, is not suited for IoT networks characterized by high density of devices and sporadic traffic of small data units. Indeed, grant-based access entails an increased protocol overhead, inducing latency and possibly network overload. By eliminating the handshake procedure, \textit{grant-free random access} was proposed as a potential solution for these problems \cite{grant_free_m2m_mimo}. Many works have proposed to combine grant-free based access and NOMA. For example, power-based grant-free NOMA was proposed in \cite{aloha_noma} using slotted ALOHA, and code-based NOMA was presented in \cite{uplink_scma} using sparse codes. We refer to \cite{grantfree_noma_survey} for a comprehensive survey on grant-free NOMA techniques. In contrast to previous works, this work aims to study the benefits of grant-free NOMA in the context of satellites and terrestrial two-hop networks with coexisting IoT services.}
\section{System model and Performance metrics}
\label{sec:system_model_performance_metrics}
\subsection{System Model}
We first consider the system illustrated in Fig.~\ref{fig:system_model}, in which $L$ APs, e.g., LEO satellites, provide connectivity to IoT devices. The APs are in turn connected to a BS, e.g., a ground station, through a shared wireless backhaul channel. \textcolor{black}{The main motivation to use the two-hop architecture in Fig. 1 is to benefit from enhanced coverage and space diversity, owing to the presence of multiple APs in the vicinity of each IoT device.} We assume that time over both access and backhaul channels is divided into frames and each frame contains $T$ time slots. At the beginning of each frame, a random number of IoT devices are active. The number of active IoT devices that generate \ac{CS} and \ac{NCS} messages at the beginning of the frame follow independent Poisson distributions with average loads $\gamma_c G$ and $(1-\gamma_c) G\ \mathrm{[packet/frame]}$, respectively, for some parameter $\gamma_c \in [0,1]$ and total load $G$. Users select a time-slot uniformly at random among the $T$ time-slots in the frame and independently from each other. By the Poisson thinning property \cite{billingsley2008probability}, the random number $N_c(t)$ of \ac{CS} messages transmitted in a time-slot $t$ follows a Poisson distribution with average $G_c = \gamma_c G /T \ \mathrm{[packet/slot]}$, while the random number $N_{\bar{c}}(t)$ of \ac{NCS} messages transmitted in slot $t$ follows a Poisson distribution with average $G_{\Bar{c}}=(1-\gamma_c)G/T \ \mathrm{[packet/slot]}$.\par
\textit{Radio Access Model:} As in, e.g., \cite{frederico2019modern,azimi2017content,calderbank_erasure}, we model the access links between any device and an AP as an independent interfering erasure channel with erasure probability $\epsilon_1$. In satellite applications, as represented in Fig.~\ref{fig:system_model_space}, this captures the presence or absence of a line-of-sight link between the transmitter and the receiver. A packet sent by a user is independently erased at each receiver with probability $\epsilon_1$, causing no interference, or is received with full power with probability $1-\epsilon_1$. The erasure channels are independent and identically distributed (i.i.d.) across all slots and frames. Interference from messages of the same type received at an AP is assumed to cause a destructive collision. Furthermore, \ac{CS} messages are assumed to be transmitted with a higher power than \ac{NCS} messages so as to improve their packet success rate, hence creating significant interference on \ac{NCS} messages. As a result, in each time-slot, an AP can be in three possible states:
\begin{itemize}
    \item a \ac{CS} message is retrieved successfully if the AP receives only one (non-erased) \ac{CS} message and no more than a number $K$ of (non-erased) \ac{NCS} messages. This implies that, due to their lower transmission power, \ac{NCS} messages generate a tolerable level of interference on \ac{CS} messages as long as their number does not exceed the threshold $K$;
    \item a \ac{NCS} message is retrieved successfully if the AP receives only one (non-erased) \ac{NCS} message;
    \item no message is retrieved otherwise.
\end{itemize}
We note that packet re-transmissions are ignored due to latency and low power constraints.
\par
\textit{Backhaul model:} The APs share a wireless out-of-band backhaul that operates in a full-duplex mode and in an uncoordinated fashion as in \cite{frederico2019modern}. The lack of coordination among APs can be considered as a worst-case scenario in dense low-cost terrestrial cellular deployments \cite{het_networks_no_coordination} \cite{vladimir_cooperative_ALOHA} and as the standard solution for constellations of LEO satellites that act as relays between ground terminals and a central ground station. In fact, satellite coordination, although feasible through the use of inter-satellite links \cite{inter_satellite_links}, may be costly in terms of on-board resources. In each time-slot $t+1$, an AP sends a message retrieved on the radio access channel in the corresponding slot $t$ over the backhaul channel to the BS. APs with no message retrieved in slot $t$ remain silent in the corresponding backhaul slot $t+1$. The link between each AP and the BS is modeled as an erasure channel with erasure probability $\epsilon_2$, and destructive collisions occur at the BS if two or more messages of the same type are received. As for the radio access case, erasure channels are i.i.d. across APs, slots and frames. We note that different forwarding strategies and buffering can be modelled by defining a forwarding probability at each AP. This can easily be accounted for in our model by simply multiplying the probability of successfully receiving a packet at each AP by the forwarding probability. We refer to \cite{munari2019multiple_relays_SA} for details.
\textcolor{black}{Due to the sporadicity and unpredictability of IoT traffic, grant-based access is problematic as it induces additional latency and under-utilization of radio resources. In this paper, we assume that radio and backhaul access is carried out using grant-free non-orthogonal slotted ALOHA \cite{RA_lte_noma}}. \textcolor{black}{Furthermore, We assume that, over both the access and backhaul channels, a simple power-based non-orthogonal multiple access scheme is used whereby the packet received with the strongest power can be decoded. For practical considerations, we do not implement successive interference cancellation by exploiting the coherent sum of different message copies.} \par
In order to model interference between APs, we consider two scenarios. The first, referred to as \textit{collision model}, assumes that multiple messages from the same or distinct device cause destructive collision. Under this model, in each time-slot, the BS's receiver can be in three possible states:
\begin{itemize}
    \item as for radio access, a \ac{CS} message is retrieved successfully at the BS if only one \ac{CS} message is received from any AP, along with no more than $K$ \ac{NCS} messages; 
    \item a \ac{NCS} message is retrieved successfully if no other \ac{CS} or \ac{NCS} message is received;
    \item no message is retrieved at the BS otherwise.
\end{itemize}
In the second model, referred to as \textit{superposition model}, the BS is able to decode from the superposition of multiple instances of the same packet that are relayed by different APs on the same backhaul slot, assuming no collisions from other transmissions. \textcolor{black}{The superposition model is motivated by the fact that, when multiple copies of the same message are received by a receiver over the same radio resource, they undergo superposition over the wireless channel. This results in an effective channel given by the sum of the channels affecting each copy of the message. Therefore, the message can be decoded without the need for interference cancellation.} In practice, this can be accomplished by ensuring that the time asynchronism between APs is no larger that the cyclic prefix in a multicarrier modulation implementation. Synchronization can be ensured, for example, by having a central master clock at the BS against which the local time bases of APs are synchronized \cite{timesynchro_patent_AP}. Overall, the BS's receiver can be in three possible states:
\begin{itemize}
    \item a \ac{CS} message is retrieved successfully at the BS in a given time-slot if no other \ac{CS} message and no more than $K$ \ac{NCS} messages are received by the BS;
    \item a \ac{NCS} message is retrieved successfully if no \ac{CS} messages and no other \ac{NCS} messages are received in the same slot;
    \item no message is retrieved at the BS otherwise.
\end{itemize}
\textcolor{black}{Note that the main difference between the collision and superposition models is that, under the collision model, a packet can not be recovered when multiple instances of the same packet are received at the BS, while this same packet can be recovered under the superposition model.}
\par \textit{Inter-service TDMA: }In addition to non-orthogonal resource allocation whereby devices from both services share the entire frame of $T$ time-slots, we also consider orthogonal resource allocation, namely \textit{inter-service time division multiple access} (TDMA), whereby a fraction $\alpha T$ of the frame's time-slots are reserved to \ac{CS} devices and the remaining $(1-\alpha)T$ for \ac{NCS} devices. Inter-service contention in each allocated fraction follows a \ac{SA} protocol as discussed above. In the following, we derive the performance metrics under the more general non-orthogonal scheme described above. The performance metrics under TDMA for each service can be directly obtained by replacing $T$ with the corresponding fraction of resources in the performance metrics equations and setting the interference from the other service to zero.
\subsection{Performance Metrics}
\label{sec:performance_metrics}
We are interested in computing the throughput $R_c$ and $R_{\bar{c}}\ [\mathrm{packet/slot}]$ and the packet success rate $\Gamma_c$ and $\Gamma_{\bar{c}}\ [\mathrm{packet/frame}]$ for \ac{CS} and \ac{NCS} respectively. The throughput is defined as the average number of packets received correctly in any given time-slot at the BS for each type of service. The packet success rate is defined as the average probability of successful transmission of a given user given that the user is active, i.e., that it transmits a packet in a given frame.
\subsection{Considerations on the System Model Assumptions}
Before moving to the analysis, some remarks on the considered system model are in order.
First, the binary erasure collision channel model with i.i.d. erasures-also referred to as on-off fading model in the literature \cite{OnOff2003}-entails some simplifications that need to be discussed. When the aggregate channel traffic is high, in a realistic scenario, the aggregate interference power, even if very small on a single packet level, may become relevant and hinder the correct reception of a data unit when many concurrent packets are transmitted. This particular aspect is not captured by the considered channel model, which only models independent erasures. However, the relevance of this scenario remains limited since high channel load conditions in an \ac{SA}-based system yield unacceptable packet success rates \cite{Munari15:Massap}. From this standpoint, the model we consider captures well the behaviour of practical systems in the more interesting low-to-moderate channel load conditions.\par
\textcolor{black}{Thirdly, it is important to note that our model accounts for various types of collisions among packets, namely, inter-CS and inter-NCS collisions, CS to NCA collisions and finally NCS to CS collisions.  }\par
Secondly, assuming independent erasures across all links provides an upper bound on the achievable performance in more realistic configurations in which fading events can render the link towards multiple relays correlated \cite{Formaggio20:CorrEr}. In general, this bound is expected to provide a good approximation for LEO constellation links towards different satellites due to the spatial separation of the channels. In contrast, this assumption may be an optimistic estimate of the performance for terrestrial systems with denser deployments. \par 
Finally, the use of uncoordinated access in the APs-to-BS links should be considered against the use of coordinated-access mechanisms. While the latter can ideally provide higher throughputs, it may also become inefficient when the traffic activity seen at the APs varies heavily, is difficult to predict, or when APs move as in a LEO constellation. In such cases, the overhead required to coordinate the access among relays overcomes the increase in data delivery and thus uncoordinated access may become preferable.

\section{Single Service Under Collision Model}
\label{sec:oneclass}
\subsection{Performance Analysis}
\begin{sloppypar}
We start by considering the baseline case of a single service under the collision model. While this can account for either \ac{CS} or \ac{NCS}, we consider here without loss of generality only the \ac{CS} by setting $\gamma_c = 1$. We note that this model is similar to a multiple-relay \ac{SA} often referred to as modern random access protocol \cite{frederico2019modern}. An AP successfully retrieves a packet when \textit{only one} of the $N_c = n_c$ 
transmitted packets arrives unerased, i.e. with probability 
\begin{equation}
    {\psU = \nTx \,(1-\perasU) \, \perasU^{\nTx-1}}, \label{eq:p_nc}
\end{equation}
where the term $\epsilon_1^{n_c-1}$ is the probability that the remaining $n_c-1$ packets are erased. Removing the conditioning on $\NTx$, one can obtain the average radio access throughput as
\begin{align}
\mathbb{E}_{N_c}[p_{n_c}] = \sum_{\nTx=0}^{\infty} \frac{\load^\nTx \, e^{-\load}}{\nTx!} \cdot  \psU = \load (1-\perasU) \, e^{-\load (1- \perasU)},
\label{eq:truSA}
\end{align}
which corresponds to the throughput of a \ac{SA} link with erasures.
\end{sloppypar}
The overall throughput $R_c$ depends also on the backhaul channel. In particular, for a successful packet transmission, an AP must successfully decode one packet, which should then reach the BS unerased over the backhaul channel. This occurs with probability
\begin{equation}
\psD = \psU  \,(1-\perasD). \label{eq:q_nc_single_Service}   
\end{equation}
In addition, the packet should not collide with other packets. By virtue of the independence of erasure events, the number of incoming packets on the backhaul during a slot follows the binomial distribution $\mathrm{Bin}(\nRx,\psD)$. Recalling that collisions are regarded as destructive under the collision model, a packet is retrieved only when a single packet reaches the BS, i.e. with probability \mbox{$q_c = \nRx \psD (1-\psD)^{\nRx-1}$}. The \ac{CS} throughput can then be derived as
\begin{equation}
R_c = \mathbb{E}_{N_c}[q_c] = \sum_{\nTx=0}^{\infty} \frac{\load^\nTx e^{-\load}}{\nTx!} \cdot \nRx \, \psD (1-\psD)^{\nRx-1}.
\label{eq:truSum}
\end{equation}
This can be computed in closed form as stated in the following proposition.\par
\textit{\textbf{Proposition 1:} Under the collision model, assuming $\gamma_c = 1$ (single service), the throughput $R_c$ is given as function of the number of APs $L$, channel erasure probabilities $\epsilon_1$ and $\epsilon_2$, and \ac{CS} packet load $G_c$ as}
\begin{equation}
\begin{aligned}
R_c = \sum_{\ell=0}^{\nRx -1} (-1)^\ell \, \nRx \, {\nRx-1 \choose \ell}& \left[\frac{(1-\perasU) (1-\perasD)}{\perasU}\right]^{\ell+1} \\ &\cdot e^{-\load} \cdot \ancF_{\ell+1}\!\left(\load \,\perasU^{\ell+1}\right),
 \label{eq:rate_single_service}
\end{aligned}
\end{equation}
\textit{where the auxiliary function $\ancF_m(x)$ is defined recursively as
}\begin{align}
\begin{split}
\ancF_0(x) &= e^x\\
\mathrm{and}\ \ancF_m(x) & = x \sum_{\ell=0}^{m-1} {m-1 \choose \ell} \ancF_\ell (x) \quad m\geq1 .
\end{split}
\label{eq:ancFunc}
\end{align} \par
\textbf{\textit{Proof: }}
Denoting  $\beta =  (1-\perasU) (1-\perasD)$, and recalling the definitions of probabilities $\psU$ and $\psD$ in equations \eqref{eq:p_nc} and \eqref{eq:q_nc_single_Service}, the throughput \eqref{eq:truSum} can be written as
\begin{align}
\begin{split}
R_c &= \sum_{\nTx=0}^{\infty} \frac{\load^\nTx e^{-\load}}{\nTx!} \cdot \nRx \, \beta \, \nTx \perasU^{\nTx-1} \left( 1 - \beta \, \nTx \perasU^{\nTx-1}\right)^{\nRx-1}\\
&\stackrel{(a)}{=}\sum_{i=0}^{\nRx-1}(-1)^i \, \nRx  {\nRx-1 \choose i} \frac{\beta^{i+1} \, e^{-\load}}{\perasU^{i+1}} \\ & \hspace{25 mm}\cdot  \sum_{\nTx=0}^{\infty} \frac{\left(\load \, \perasU^{i+1}\right)^\nTx}{\nTx!} \cdot \nTx^{i+1},
\end{split}
\label{eq:truDerivation}
\end{align}
where $(a)$ follows by applying Newton's binomial expansion and after some simple yet tedious rearrangements. Let us now introduce the auxiliary function
\begin{equation}
\ancF_{m}(x) = \sum_{\nTx=0}^{\infty} \frac{x^\nTx \,\nTx^m}{\nTx!}.
\end{equation}
From the definition of Taylor's series for the exponential function, we have $\ancF_0(x) = e^x$. Moreover, for $m\geq1$, we have
\begin{align}
\ancF_m(x) &= x \sum_{\nTx=0}^{\infty} \frac{x^{\nTx-1} \,\nTx^{m-1}}{(\nTx-1)!} \stackrel{(b)}{=} \,\,x \sum_{t=0}^{\infty} \frac{x^{t} \,(t+1)^{m-1}}{t!} \\
&\stackrel{(c)}{=} x \sum_{\ell=0}^{m-1} {m-1 \choose \ell}\sum_{t=0}^{\infty} \frac{x^{t} \,t^{\ell}}{t!}\\
&= x \sum_{\ell=0}^{m-1} {m-1 \choose \ell}\ancF_\ell(x),
\end{align}
where equality $(b)$ applies the change of variable $t=\nTx-1$ and equality $(c)$ results from applying once more Newton's binomial expansion to $(t+1)^{m-1}$. Plugging this result into the innermost summation within \eqref{eq:truDerivation} leads to the closed form expression of the \ac{CS} throughput reported in \eqref{eq:rate_single_service}. \qed
\begin{figure}[t]
	\centering
	\includegraphics[height= 6.3 cm, width= 8.5 cm]{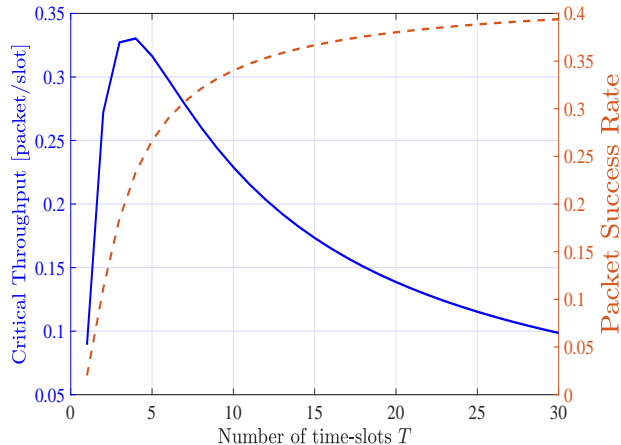}
	\caption{Single service (here \ac{CS}) throughput (solid line) and packet success rate (dashed line) as a function of the number of time-slots $T$ ($\epsilon = 0.5$, $G=16\ [\mathrm{packet/frame}]$, $\gamma_c = 1$ and $L = 3$ APs).}
	\label{fig:single_service}
\end{figure}
\par
We now turn to the packet success rate. Define the RV $N^{\prime}_c \geq 1$ to count the number of transmitted messages given that at least one message is transmitted. This RV has the distribution
\begin{equation}
    P(N^{\prime}_c = n^{\prime}_c | N^{\prime}_c \geq 1) = (1-e^{-G_c})^{-1} \cdot  \frac{e^{-G_c}G_c^{n^{\prime}_c}}{n^{\prime}_c!} \label{eq:n_primec_distribution}
\end{equation}
which corresponds to a normalized Poisson distribution over the set $\{1, \dots, \infty \}$. For a given value $N^{\prime}_{c} = n^{\prime}_{c}$, the probability that the packet of a given user $u$ reaches an AP given that $u$ is active is given by $p_u = (1-\epsilon_1)\epsilon_1^{n^{\prime}_c-1}$. Furthermore, the probability that the user's packet reaches the BS is given as
\begin{equation}
    q_u = \underbrace{L p_u (1-\epsilon_2)}_{(a)} \underbrace{(1-q_{n^{\prime}_{c}})^{L-1}}_{(b)}, \label{eq:q_u_collision_single_class}
\end{equation}
where $p_{n^{\prime}_{c}}$ and $q_{n^{\prime}_{c}}$ are defined in \eqref{eq:p_nc} and \eqref{eq:q_nc_single_Service}. In \eqref{eq:q_u_collision_single_class}, term $(a)$ is the probability that the user's packet is received at the BS from any of the $L$ APs, while $(b)$ denotes the probability that the BS does not receive any \ac{CS} message from the remaining $L-1$ APs. The packet success rate $\Gamma_c$, can be obtained by averaging \eqref{eq:q_u_collision_single_class} over $N^{\prime}_c$ as $\Gamma_c = \mathbb{E}_{N^{\prime}_c}[q_u]$. This can be obtained in closed form as stated in the following proposition.\par
\textit{\textbf{Proposition 2:} Under the collision model, assuming $\gamma_c = 1$ (single service), the packet success rate $\Gamma_c$ is given as function of the number of APs $L$, channel erasure probabilities $\epsilon_1$ and $\epsilon_2$, and \ac{CS} packet load $G_c$ as}
\begin{equation}
\begin{aligned}
    \Gamma_c = L\beta(1-&e^{-G_c})^{-1}e^{-G_c} \sum_{l=0}^{L-1} \frac{(-\beta)^l}{\epsilon_{1}^{l+1}} {L-1 \choose l}\\ & \cdot \Big[\mathbbm{1}_{l=0}(e^{\epsilon_1 G_c}-1) + \mathbbm{1}_{l>0} \mathcal{H}_l (G_c \epsilon_1^{l+1}) \Big], \label{eq:loss_single_service}
    \end{aligned}
\end{equation}
\textit{where the function $\mathcal{H}_l (\cdot)$ is defined in \eqref{eq:ancFunc} and we have $\beta = (1-\epsilon_1)(1-\epsilon_2)$}.
\par
\textbf{\textit{Proof: }}The proof follows using the same steps as for \textit{Proposition 1}. \qed \\
We note that in the regime of high number of APs, i.e., $L \to \infty$, the throughput and packet success rate derived in \textit{Proposition 1} and \textit{Proposition 2} are equal to zero as detailed in Appendix D. This shows that the number of APs should be carefully selected. This will be further investigated in Sec. \ref{sec:heterogeneous_superposition} for heterogeneous services.
\subsection{Examples}
Using the expressions derived in \textit{Proposition} 1 and 2, we plot in Fig.~\ref{fig:single_service} the throughput and packet success rate for a single service as function of the number of time-slots $T$. Increasing $T$ is seen to improve the packet success rate: an active user has a larger chance of successful transmission when more time-slots are available for random access. In contrast, there exist an optimal value of $T$ for the throughput, as hinted by the analysis of the standard ALOHA protocol. Increasing $T$ beyond this optimal value reduces the throughput owing to the larger number of idle time-slots. The asymptotic behaviors of packet success rate and throughput can be easily verified theoretically using the expressions in \textit{Proposition 1} and \textit{Proposition 2} by taking their limit when $G_c$ tends to zero.
\section{Heterogeneous Services Under Collision Model}
\label{sec:heterogeneous_collision}
In this section, we extend the analysis in the previous section to derive the throughput and packet success rate of both CS and NCS under the collision model described in Sec. \ref{sec:system_model_performance_metrics}.
\subsection{Heterogeneous Services with Ideal NCS-to-CS Interference Tolerance}
\label{sec:het_K_infty}
We start by considering the case in which decoding of \ac{CS} messages is not affected by \ac{NCS} traffic, i.e., we set $K \to \infty$. Under this assumption, the \ac{CS} throughput and packet success rate expressions equals the expressions in Propositions 1 and 2. We hence focus here on the performance of \ac{NCS}, as summarized in the following proposition.\par
\textit{\textbf{Proposition 3:} Under the collision model with ideal \ac{NCS}-to-\ac{CS} interference tolerance, i.e. $K \to \infty$, the \ac{NCS} throughput $R_{\bar{c}}$ and packet success rate $\Gamma_{\bar{c}}$ can be respectively written as a function of the number of APs $L$, channel erasure probabilities $\epsilon_1$ and $\epsilon_2$, and \ac{CS} and \ac{NCS} packet loads $G_c$ and $G_{\bar{c}}$ as }
\begin{align}
\begin{split}
R_{\bar{c}} = L \sum_{i=0}^{L-1} \sum_{k=0}^{i} (-1)^i &  {L-1 \choose i}{i \choose k}  \Bigg( \frac{\beta}{\epsilon_1} \Bigg)^{i+1} e^{-G} \\ & \cdot
  \mathcal{H}_{i-k}(G_c\epsilon_1^{i + 1}) \mathcal{H}_{k+1}(G_{\bar{c}}\epsilon_1^{k+1})
 \label{eq:R_barc_no_K_collision}
\end{split}
\end{align}
\begin{equation}
    \begin{aligned}
    \mathrm{and}\ \Gamma_{\bar{c}} & = \sum_{l=0}^{L-1} L \beta^{l+1} \epsilon_1^{-(l+1)} (-1)^{-l} (1-e^{-G_{\bar{c}}})^{-1} \\ & \cdot {L-1 \choose l} \bigg\{ \sum_{m=0}^{l} {l \choose m} A \cdot B  + \\ & e^{-G} [\mathbbm{1}_{l=0} (e^{\epsilon_1 G_{\bar{c}}}-1) + \mathbbm{1}_{l>0} \mathcal{H}_{l}(G_{\bar{c}} \epsilon_1^{l+1})]\bigg\} , \label{eq:loss_barc_no_K_collision}
    \end{aligned}
\end{equation}
\textit{where} 
\begin{equation}
    \begin{aligned}
    A = e^{-G_c} (e^{G_c \epsilon_1^{l+1}}-1) \mathbbm{1}_{m=0} +  e^{-G_c} \mathcal{H}_{m}(G_c \epsilon_1^{l+1}) \mathbbm{1}_{m > 0}
    \end{aligned}
\end{equation}
\begin{equation}
\begin{aligned}
 B = e^{-G_{\bar{c}}} \mathcal{H}_{l-m}(G_{\bar{c}} \epsilon_1^{l-m+1}) & \mathbbm{1}_{l-m \neq 0} \\ & + e^{-G_{\bar{c}}}(e^{\epsilon_1G_{\bar{c}}} - 1) \mathbbm{1}_{l-m = 0}.
 \end{aligned}
\end{equation}
\textbf{\textit{Proof: }} The proof is detailed in Appendix A. \qed
\subsection{Examples}
\label{sec:num_results_het_coll}
\begin{figure}[t]
	\centering
	\includegraphics[height= 6.3 cm, width= 9.5 cm]{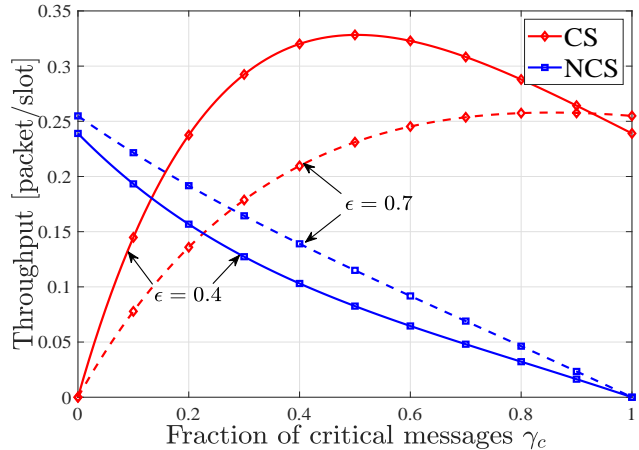}
	\caption{\ac{CS} and \ac{NCS} throughput as function of the fraction of \ac{CS} messages $\gamma_c$ for the collision model and using non-orthogonal resource allocation ($\epsilon_1=\epsilon_2=\epsilon=0.4$ or $0.7$, $G=8\ [\mathrm{packet/frame}]$, $T=4\ [\mathrm{time\textrm{-}slot/frame}]$, and $L = 3$ APs).}
	\label{fig:rate_load_1}
\end{figure}

In order to study the performance trade-offs between the two services, we start by investigating the impact of $\gamma_c$ by plotting in Fig. \ref{fig:rate_load_1} the \ac{CS} and \ac{NCS} throughputs versus $\gamma_c$ with $\epsilon_1=\epsilon_2=\epsilon$, $G=30\ [\mathrm{packet/frame}]$, $T=4\ [\mathrm{time\text{-}slot/frame}]$, and $L = 3$ APs. For \ac{CS}, there is an optimal value of $\gamma_c$ that ensures an optimized \ac{CS} load as in the standard analysis of the ALOHA protocol, discussed also in the context of Fig. \ref{fig:single_service}. In contrast, the \ac{NCS} throughput decreases as function of $\gamma_c$ due to the increasing interference from \ac{CS} transmissions. The \ac{NCS} throughput is also seen to increase as a function of the channel erasure $\epsilon$ when $\epsilon$ is not too large. This is because a larger $\epsilon$ can reduce the interference from $\ac{CS}$ transmissions.

\subsection{Heterogeneous Services with Limited NCS-to-CS Interference Tolerance}
We now alleviate the assumption that \ac{CS} transmissions can withstand any level of \ac{NCS} interference by assuming that interference from at most $K$ \ac{NCS} transmissions can be tolerated without causing a collision from \ac{CS} traffic. We derive both \ac{CS} and \ac{NCS} performance metrics. We note that, perhaps counter-intuitively, both \ac{CS} and \ac{NCS} performance metrics are affected by the \ac{CS} interference tolerance parameter $K$. In fact, with a lower value of $K$, a smaller number of \ac{CS} packets tends to reach the BS, reducing interference to \ac{NCS} transmissions. We start by detailing the \ac{NCS} performance metrics.\par
\textbf{\textit{Proposition 4: }}\textit{Under the collision model with limited \ac{NCS}-to-\ac{CS} interference tolerance, i.e. finite $K$, the \ac{NCS} throughput and packet success rate are given as a function of the number of APs $L$, channel erasure probabilities $\epsilon_1$ and $\epsilon_2$, and \ac{CS} and \ac{NCS} packet loads $G_c$ and $G_{\bar{c}}$ as}
\begin{equation}
\begin{aligned}
  R_{\bar{c}} =&  L \sum_{i=0}^{L-1} \sum_{k=0}^{i} (-1)^i   {L-1 \choose i}{i \choose k} \Bigg( \frac{\beta}{\epsilon_1} \Bigg)^{i+1} \\ & \cdot e^{-G} 
  \mathcal{H}_{i-k}(G_c\epsilon_1^{i + 1})\Big[ \xi_1(K, L, G_{\bar{c}}, \epsilon_1) + \xi_2(K, L, G_{\bar{c}}, \epsilon_1) \Big] \label{eq:R_barc_with_K_collision}
  \end{aligned}
\end{equation}
\begin{equation}
    \mathrm{and}\ \Gamma_{\bar{c}} = \mathbb{E}_{N_c, N^{\prime}_{\bar{c}}}\bigg[  L (1-\epsilon_1)\epsilon_1^{N_{\bar{c}}^{\prime}-1} \epsilon_1^{N_c} (1-\epsilon_2)(1-q)^{L-1}\bigg], \label{eq:gamma_barc_with_K_collision}
\end{equation}
\textit{where $q = (1-\epsilon_2)(p_{N_c} \gamma_{K-1}(N_{\bar{c}}^{\prime}, \epsilon_1) + N_{\bar{c}}^{\prime}(1-\epsilon_1)\epsilon_1^{N^{\prime}_{\bar{c}}-1} \epsilon_1^{N_c})$,}
\begin{subequations}
\begin{alignat}{1}
& \xi_1(K, L, G_{\bar{c}}, \epsilon_1) = \sum_{n_{\bar{c}}=0}^{K} \frac{(G_{\bar{c}} \epsilon_1^{k+1})^{n_{\bar{c}}}}{n_{\bar{c}}!} n_{\bar{c}}^{k+1} \\
& \xi_2(K, L, G_{\bar{c}}, \epsilon_1) =  \sum_{n_{\bar{c}}=K+1}^{+\infty} \frac{(G_{\bar{c}} \epsilon_1^{k+1})^{n_{\bar{c}}}}{n_{\bar{c}}!}\\& \hspace{15 mm}\cdot n_{\bar{c}}^{k+1} \bigg[\sum_{l=0}^{K} {n_{\bar{c}} \choose l} (1-\epsilon_1)^{l} \epsilon_1^{n_{\bar{c}}-l}\bigg]^{i-k},
\end{alignat}
\end{subequations}
\textit{and the expectation in \eqref{eq:gamma_barc_with_K_collision} is taken with respect to independent RVs $N_c$ and $N^{\prime}_{\bar{c}}$, with the latter distributed as in \eqref{eq:n_primec_distribution} with the index $\bar{c}$ in lieu of $c$.}
\par
\textbf{\textit{Proof: }}The proof is detailed in Appendix D. \qed \par
We now address the \ac{CS} analysis.
With finite $K$, a \ac{CS} message is correctly received at any AP if it is the only non-erased \ac{CS} message and no more than $K$ \ac{NCS} messages are received erasure-free at the AP. Conditioned on the number of messages $N_c=n_c$ and $N_{\bar{c}} = n_{\bar{c}}$, the probability of the first event is given by $p_{n_c}$ defined in \eqref{eq:p_nc}, while the probability of the second event is given by $\gamma_K(n_{\bar{c}}, \epsilon_1)$ with
\begin{equation}
    \gamma_K(x, \epsilon)  = \begin{cases}
    1 & \text{if}\quad x \leq K  \\
    \sum_{i=0}^{K} {x \choose i} \left(1-\epsilon \right)^i \epsilon^{x - i} &  \text{otherwise.}
    \end{cases} \label{eq:gamma_K}
\end{equation}
Removing the conditioning on $N_{\bar{c}}$, the probability of the second event can be written as 
\begin{align}
\begin{split}
& \sum_{\nTxNC=0}^{K} \frac{\loadNC^\nTxNC \, e^{-\loadNC}}{\nTxNC!}+\!\!\!\!\!\!\! \sum_{\nTxNC=K+1}^{\infty} \!\!\!\frac{\loadNC^\nTxNC \, e^{-\loadNC}}{\nTxNC!}\left[ \sum_{i=0}^{K} {\nTxNC \choose i} \left(1-\perasU \right)^i \perasU^{\nTxNC - i}\right] \\
& = Q(K+1,\loadNC)+\xi(K,\loadNC),
\end{split}
\label{eq:gamma_K_no_conditioning}
\end{align}
where the first term in \eqref{eq:gamma_K_no_conditioning} is the regularized gamma function and $\xi(K,\loadNC)$ represents the second term. For the \ac{CS} performance metrics we distinguish the following two cases depending on the number $L$ of APs. \par
\subsubsection{Small number of APs ($\nRx\leq K+1$)}In this case, the effect of finite interference tolerance $K$ affects only the radio access transmission phase. In fact, in the backhaul transmission phase, if $\nRx\leq K+1$, the number of interfering \ac{NCS} transmissions on a \ac{CS} packet at the BS cannot exceed $K$.\par
\textit{\textbf{Proposition 5:} Under the collision model, the \ac{CS} throughput and packet success rate given as a function of the number of APs $\nRx\leq K+1$, channel erasure probabilities $\epsilon_1$ and $\epsilon_2$ and \ac{CS} and \ac{NCS} packet loads $G_c$ and $G_{\bar{c}}$ as}
\begin{align}
\begin{split}
R_c &= \sum_{\ell=0}^{\nRx -1} (-1)^\ell \, \nRx \, {\nRx-1 \choose \ell} \left[\frac{ (1-\perasU) (1-\perasD)}{\perasU}\right]^{\ell+1} \\ & \cdot e^{-\load} \,\,\ancF_{\ell+1}\left(\load \,\perasU^{\ell+1}\right)\cdot \left[\mathsf{Q}(K+1,\loadNC)+\xi(K,\loadNC,\ell)\right]
\end{split}
\label{eq:truDKfinite}
\end{align}
\begin{align}
\mathrm{and}\ \ \Gamma_c = \mathbb{E}_{N^{\prime}_c, N_{\bar{c}}}[L p_u (1-\epsilon_2)(1-q_{N^{\prime}_{c}})^{L-1}],
\label{eq:reliabilityKfinite}
\end{align}
\textit{where}
\begin{equation}
    \xi(K,\loadNC,\ell)=\!\!\sum_{\nTxNC=K+1}^{\infty} \!\!\frac{\loadNC^\nTxNC \, e^{-\loadNC}}{\nTxNC!}\!\left[ \sum_{i=0}^{K} {\nTxNC \choose i} \left(1-\perasU \right)^i \!\perasU^{\nTxNC - i}\!\right]^{\!\ell\!+\!1}
\end{equation}\textit{and} $q_{n^{\prime}_c} = n^{\prime}_c (1-\epsilon_1) \epsilon_1^{n^{\prime}_{c}-1}\gamma_K(n_{\bar{c}}, \epsilon_1) (1-\epsilon_2)$ \textit{is the probability of receiving any \ac{CS} packet at the BS. 
}
\textit{and the expectation in \eqref{eq:reliabilityKfinite} is taken with respect to independent RVs $N_{\bar{c}}$ and $N^{\prime}_c$, with the latter distributed as in \eqref{eq:n_primec_distribution}}.\\
\textit{\textbf{Proof: }}The proof is provided in Appendix E. \qed \\ 
Comparing the \ac{CS} throughput in \eqref{eq:truDKfinite} with the expression \eqref{eq:rate_single_service} for $K\rightarrow \infty$ we observe that the effect of a finite interference tolerance is measured by the multiplicative term ${\left[Q(K+1,\loadNC)+\xi(K,\loadNC,\ell)\right]}$. It can be shown that this term is always smaller than one, which is in line with the fact that a lower \ac{CS} throughput is expected when $K$ is finite.
\subsubsection{Large Number of APs ($\nRx> K+1$)}In this case, a successful \ac{CS} transmission occurs in all events where a single \ac{CS} packet and only up to $K<\nRx$ \ac{NCS} packets reach the BS. The total probability of these events given $N_c = n_c$ and $N_{\bar{c}} = n_{\bar{c}}$ can be computed as
\begin{equation}
q_c= \sum_{\ell=0}^{K} {\nRx \choose {1,\ell,\nRx-\ell-1}}q^{\prime}_{n_c} (q_{n_{\bar{c}}})^\ell(1-q^{\prime}_{n_c}-q_{n_{\bar{c}}})^{\nRx-\ell-1},
\label{eq:truDKfiniteSum_2}
\end{equation}
where $q^{\prime}_{n_c}=q_{n_{\bar{c}}}\gamma_K(n_{\bar{c}}, \epsilon_1)\,(1-\perasD)$ is the probability that a \ac{CS} packet reaches the BS and $q_{n_{\bar{c}}}$ is the probability that a \ac{NCS} packet reaches the BS (may also be not correctly received due to a collision). Removing the conditioning on $\NTx=\nTx$ and $\NTxNC=\nTxNC$, we get the \ac{CS} throughput as
\begin{align}
\begin{split}
R_c &= \mathbb{E}_{\NTx, \NTxNC}\left[q_c\right],
\end{split}
\label{eq:truDKfiniteLGK}
\end{align}
where the expectation is taken with respect to independent RVs $N_{\bar{c}}$ and $N_c$.\par
Moving to the \ac{CS} packet success rate, conditioned on $N_c = n_c$ and $N_{\bar{c}} = n_{\bar{c}}$, the probability $p_u$ of receiving a packet at an AP from a given user $u$ is given as in $p_u = (1-\epsilon_1)\epsilon_1^{n^\prime_c-1} \gamma_K(n_{\bar{c}}, \epsilon_1)$. The probability of receiving successfully a \ac{CS} packet at the BS is then given as 
\begin{equation}
    q_c = L p_u (1-\epsilon_2)(1-q_{n^{\prime}_c})^{L-1} \underbrace{\sum_{i=0}^{K} {L-1 \choose i } p_{\bar{c}}^{i} (1-p_{\bar{c}})^{L-1-i}}_{(a)} \label{eq:q_c_2}
\end{equation}
where $q_{n^{\prime}_c} = p_{n^{\prime}_c} \gamma_K(n_{\bar{c}}, \epsilon_1) (1-\epsilon_2)$ and $p_{\bar{c}} = n_{\bar{c}}(1-\epsilon_1)\epsilon_1^{n^{\prime}_c}\epsilon_1^{n_{\bar{c}}-1} (1-\epsilon_2)$.
The main difference between \eqref{eq:q_c_2} and the probability inside the expectation in \eqref{eq:reliabilityKfinite} is the multiplication by the term $(a)$ in \eqref{eq:q_c_2} which corresponds to the probability that a number of \ac{NCS} packets lower or equal to $K$ should be received in order to be able to recover a \ac{CS} packet. Removing the conditioning on $\NTx=\nTx$ and $\NTxNC=\nTxNC$, we obtain the packet success rate as
\begin{align}
\begin{split}
\Gamma_c &= \mathbb{E}_{\NTx^{\prime}, \NTxNC}\left[q_c\right],
\end{split}
\label{eq:lossDKfiniteLGK}
\end{align}
where the expectation is taken with respect to independent RVs $N^{\prime}_{\bar{c}}$ and $N^{\prime}_c$ with the latter distributed as in \eqref{eq:n_primec_distribution}.
\subsection{Examples}
\begin{figure}[t]
	\centering
	\includegraphics[height= 6.3 cm, width= 8.5 cm]{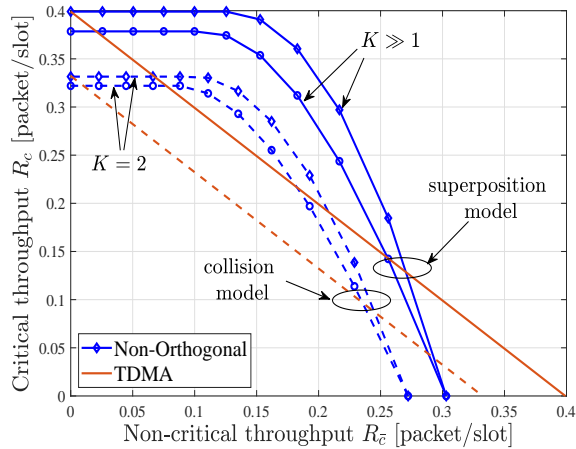}
	\caption{Achievable throughput region for \ac{CS} and \ac{NCS} under superposition and collision models for $K=2$ and $K \gg 1$ ($\epsilon = 0.5$, $G=8\ [\mathrm{packet/frame}]$, $T=2\ \mathrm{[time\textrm{-}slot/frame]}$, and $L = 3$ APs).}
	\label{fig:rate_region_K}
\end{figure}

In order to capture the effect of the number of \ac{NCS} messages $K$ on the CS, in Fig.~\ref{fig:rate_region_K} we plot the throughput region for $K=2$ and $K \to \infty$, with the latter case corresponding to the analysis in Sec.~\ref{sec:het_K_infty}. The region includes all throughput pairs that are achievable for some value of the fraction $\gamma_c$ of \ac{CS} messages, as well as all throughput pairs that are dominated by an achievable throughput pair (i.e., for which both \ac{CS} and \ac{NCS} throughputs are smaller than for an achievable pair). For reference, we also plot the throughput region for a conventional inter-service TDMA protocol, whereby a fraction $\alpha T$ for $\alpha \in [0,1]$ of the $T$ time-slots is allocated for \ac{CS} messages and the remaining time-slots to \ac{NCS} messages. For TDMA, the throughput region includes all throughput pairs that are achievable for some value of $\alpha$, as well as of $\gamma_c$.\par
A first observation from the figure is that non-orthogonal resource allocation can accommodate a significant \ac{NCS} throughput without affecting the \ac{CS} throughput, while TDMA causes a reduction in the \ac{CS} throughput for any increase in the \ac{NCS} throughput. This is due to the need in TDMA to allocate orthogonal time resources to \ac{NCS} messages in order to increase the corresponding throughput. However, with non-orthogonal resource allocation, the maximum \ac{NCS} throughput is generally penalized by the interference caused by the collisions from \ac{CS} messages, while this is not the case for TDMA. In summary, TDMA is preferable when one wishes to guarantee a large \ac{NCS} throughput and the \ac{CS} throughput requirements are loose; otherwise, non-orthogonal resource allocation outperforms TDMA in terms of throughput. Furthermore, the throughput region is generally decreased by lower value of $K$. Experiments concerning packet success rate and performance as function of the number of APs will be presented in the superposition model in the following section.
\section{Heterogeneous Services Under Superposition Model}
\label{sec:heterogeneous_superposition}
In this section, we consider the superposition model described in Sec. \ref{sec:system_model_performance_metrics}.
\subsection{Performance Analysis}
Unlike the collision model, in order to analyze the throughput and packet success rate under the superposition model, one needs to keep track of the index of the messages decoded by the APs. This is necessary to detect when multiple versions of the same message (i.e., sent by the same device) are received at the BS. Accordingly, we start by defining the RVs $B_i$ to denote the index of the message received at AP $i$ and RV $B$ for the BS at any time-slot. Accordingly, for given values $N_c=n_c$ and $N_{\bar{c}}=n_{\bar{c}}$ of transmitted messages, RVs $\{B_i\}$ can take values
\begin{equation}
    B_i = \begin{cases}
    0 & \text{if no message is retrieved}\\ & \text{due to erasures or collisions}\\
    1 \leq m \leq n_c & \text{if the $m$-th \ac{CS} message}\\&\text{is retrieved} \\
    n_c + 1 \leq m \leq n_c + n_{\bar{c}} &  \text{if the $(m-n_c)$-th \ac{NCS}}\\ & \text{ message is retrieved.}
 \label{eq:Bi_superposition}
    \end{cases}
\end{equation}
Note that we have indexed \ac{CS} messages from $1$ to $n_c$ and \ac{NCS} messages from $n_c + 1$ to $n_c + n_{\bar{c}}$. As for the RV $B$ at the BS, it is defined as 
\begin{equation}
     B = \begin{cases}
    c & \text{if a \ac{CS} message is retrieved} \\
    \bar{c} &  \text{if a \ac{NCS} message is retrieved} \\
    0 & \text{if no message is retrieved due to erasures}\\& \text{or collisions.}
    \end{cases} \label{eq:B_collision}
\end{equation}
Furthermore, we define as $M_m = \sum_{i=1}^{L} \mathbbm{1}_{\{B_i=m\}}$ the RVs denoting the number of APs that have message of index $m \in \{0,1, \ldots, n_c,n_c+1, \ldots , n_c + n_{\bar{c}} \}$. The joint distribution of RVs $\{M_m \}_{m=0}^{n_c+n_{\bar{c}}}$ given $N_c$ and $N_{\bar{c}}$ is multinomial and can be written as follows
\begin{equation}
\begin{aligned}
    &\{M_m \}_{m=0}^{n_c+n_{\bar{c}}}|N_c,N_{\bar{c}} \sim \\& \operatorname{Multinomial}\Big(L ,\overbrace{1\!-\!p_{n_c}\!-\!p_{n_{\bar{c}}}}^{0},\overbrace{\frac{p_{n_c}}{n_c}, \ldots,\frac{p_{n_c}}{n_c}}^{n_c}, \overbrace{\frac{p_{n_{\bar{c}}}}{n_{\bar{c}}},\ldots, \frac{p_{n_{\bar{c}}}}{n_{\bar{c}}}}^{n_{\bar{c}}}\Big), \label{eq:multinomial}
    \end{aligned}
\end{equation}
where we used the the probabilities in \eqref{eq:p_nc} and \eqref{eq:p_nbarc} that one of the \ac{CS} or \ac{NCS} message is received at an AP respectively in a given time-slot. 
The probability of retrieving a \ac{CS} message in a given time-slot at the BS conditioned on $N_c$, $N_{\bar{c}}$ and $\{M_{m\prime} \}_{m\prime=0}^{n_c + n_{\bar{c}}}$ can be then written as 
\begin{equation}
\begin{aligned}
    q_c&=\mathrm{Pr}[B=c | N_c=n_c , N_{\bar{c}}=n_{\bar{c}}, \{M_{m^\prime} \}_{m^\prime=0}^{n_c + n_{\bar{c}}}] \\= &\gamma_K \bigg(\sum_{m^\prime=n_c+1}^{n_{\bar{c}}+n_c} M_{m^{\prime}} , \epsilon_2 \bigg) \\ & \cdot \sum_{m=1}^{n_c} \sum_{j=1}^{M_m} {M_m \choose j} (1-\epsilon_2)^j \epsilon_2^{\sum_{ \substack{m^\prime = 0 \\ m^\prime \neq m}}^{n_c} M_{m^\prime} + M_m - j}, \label{eq:proba_c_superposition} 
    \end{aligned}
\end{equation}
where the first sum is over all possible \ac{CS} messages, the second sum is over all combinations of APs that have the \ac{CS} message $m$, and the third sum at the exponent is over all APs that have a \ac{CS} message $m^{\prime} \neq m$. The \ac{CS} throughput can be computed by averaging \eqref{eq:proba_c_superposition} over all conditioning variables as
\begin{equation}
    R_c = \mathbb{E}_{N_c , N_{\bar{c}} , \{M_m\}_{m=0}^{N_c+N_{\bar{c}}}} [q_c]. \label{eq:R_c_sup}
\end{equation}
In a similar manner, the conditional probability of receiving a \ac{NCS} message at the BS can be written as
\begin{equation}
\begin{aligned}
    &q_{\bar{c}}=\mathrm{Pr} [B = \bar{c} |  N_c=n_c , N_{\bar{c}}=n_{\bar{c}} , \{M_{m^\prime} \}_{m^\prime = 0}^{n_c + n_{\bar{c}}} ] \\ &= \sum_{m=n_c + 1}^{n_c + n_{\bar{c}}} \sum_{j=1}^{M_m} {M_m \choose j} (1-\epsilon_2)^j \epsilon_2^C,
    \end{aligned} \label{eq:proba_cbar_one}
\end{equation}
where \begin{equation}
    C = {\sum_{ \substack{m^{\prime\prime}=n_c + 1 \\ m^{\prime\prime} \neq m } }^{n_c + n_{\bar{c}}} M_{m^{\prime \prime}} + M_m - j + \sum_{m^\prime = 1}^{n_c} M_{m^\prime}} \label{eq:C}
\end{equation}
where the first sum in \eqref{eq:proba_cbar_one} is over all possible \ac{NCS} messages $m$; the second sum is over all possible combinations of APs that have message $m$. The first and second sums in \eqref{eq:C} are over all APs that have a different \ac{NCS} message and a \ac{CS} message respectively. The \ac{NCS} throughput can be then obtained by averaging over the conditioning RVs as
\begin{equation}
    R_{\bar{c}} = \mathbb{E}_{N_c, N_{\bar{c}}, \{ M_m \}_{m= 0}^{N_c + N_{\bar{c}}} } [q_{\bar{c}}]. \label{eq:R_barc_sup}
\end{equation}
\par The packet success rate under the superposition model for \ac{CS} and \ac{NCS} can be obtained by fixing $m$ to one and substituting $n_c$ and $n_{\bar{c}}$ by $n_c^{\prime}$ and $n_{\bar{c}}^{\prime}$ in \eqref{eq:proba_c_superposition} and \eqref{eq:proba_cbar_one}.
\subsection{Examples}
In Fig.~\ref{fig:rate_region_K}, we plot the throughput region for non-orthogonal resource allocation and inter-service TDMA under the superposition model. Comparing the regions of the collision model and the superposition model, it is clear that the latter provides a larger throughput region being able to leverage transmissions of the same packets from multiple APs as compared to the collision model. This can also be seen as function of $K$ in Fig. \ref{fig:rate_region_K}.
\par
\begin{figure}[h]
	\centering
	\includegraphics[height= 6.3 cm, width= 8.5 cm]{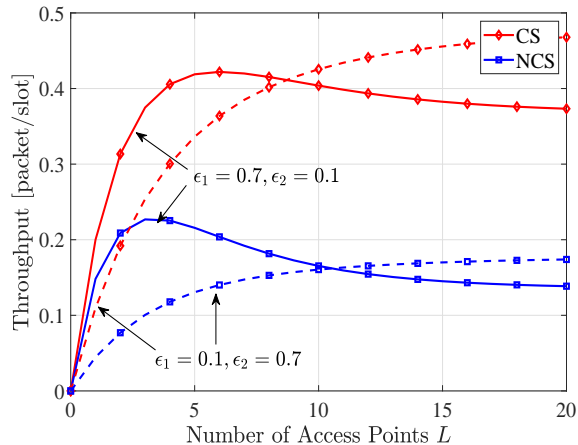}
	\caption{\ac{CS} and \ac{NCS} throughputs as function of the number of APs $L$ under the superposition model under non-orthogonal resource allocation ($G=8\ [\mathrm{packet/frame}]$, $T=4\ \mathrm{[time\textrm{-}slot/frame]}$, $\gamma_c = 0.5$ and for $\epsilon_1 \neq \epsilon_2$).}
	\label{fig:rate_L}
\end{figure}
In Fig.~\ref{fig:rate_L}, we explore the effect of the number of APs $L$ on the \ac{CS} and \ac{NCS} throughputs. In practice, in \ac{NTN}, the number of LEO satellites available can be tuned by properly designing their orbit, or by varying their speed of rotation in order to slow them down above areas of high devices density. To capture separately the effects of the radio access and the backhaul channel erasures, we consider different values for the channel erasure probabilities $\epsilon_1$ and $\epsilon_2$. We highlight two different regimes: the first is when $\epsilon_1$ is large and $\epsilon_2$ is small, and hence larger erasures occur on the access channel; while the second covers the complementary case where $\epsilon_1$ is small and $\epsilon_2$ is large. In the first regime, increasing the number of APs is initially beneficial to both \ac{CS} and \ac{NCS} messages in order to provide additional spatial diversity for the radio access, given the large value of $\epsilon_1$; but larger values of $L$ eventually increase the probability of collisions at the BS on the backhaul due to the low value of $\epsilon_2$. In the second regime, when $\epsilon_1=0.1$ and $\epsilon_2=0.8$ much lower throughputs are generally obtained due to the significant losses on the backhaul channel. This can be mitigated by increasing the number of APs, which increases the probability of receiving a packet at the BS.\par
\begin{figure*}
\centering
\begin{subfigure}{.5\textwidth}
  \centering
  \includegraphics[height=5.5 cm, width= 7.5 cm]{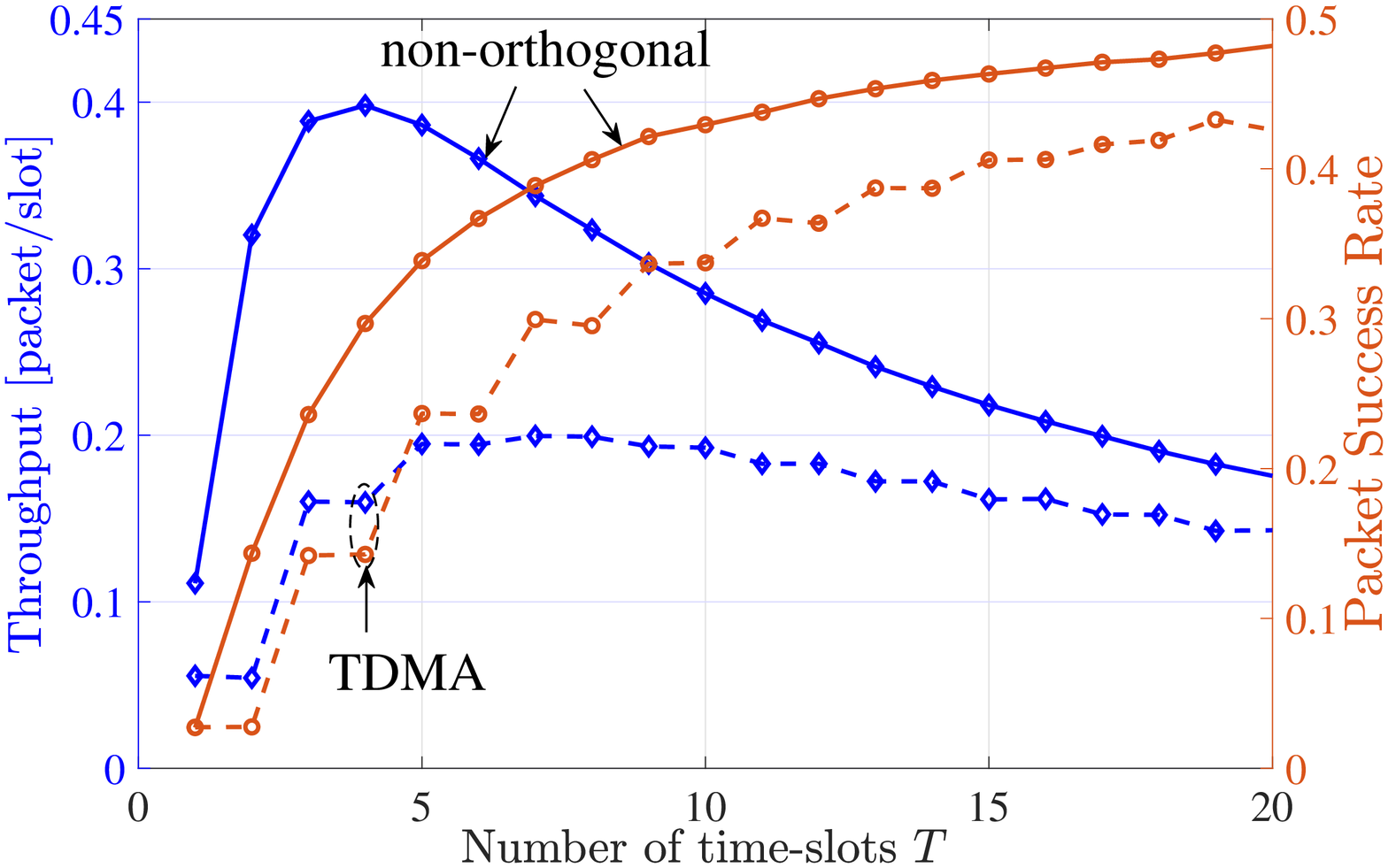}
  \caption{\ac{CS} throughput and packet success rate}
  \label{fig:throughput_reliability_critical_messages}
\end{subfigure}%
\begin{subfigure}{.5\textwidth}
  \centering
  \includegraphics[height= 5.5 cm, width= 7.5 cm]{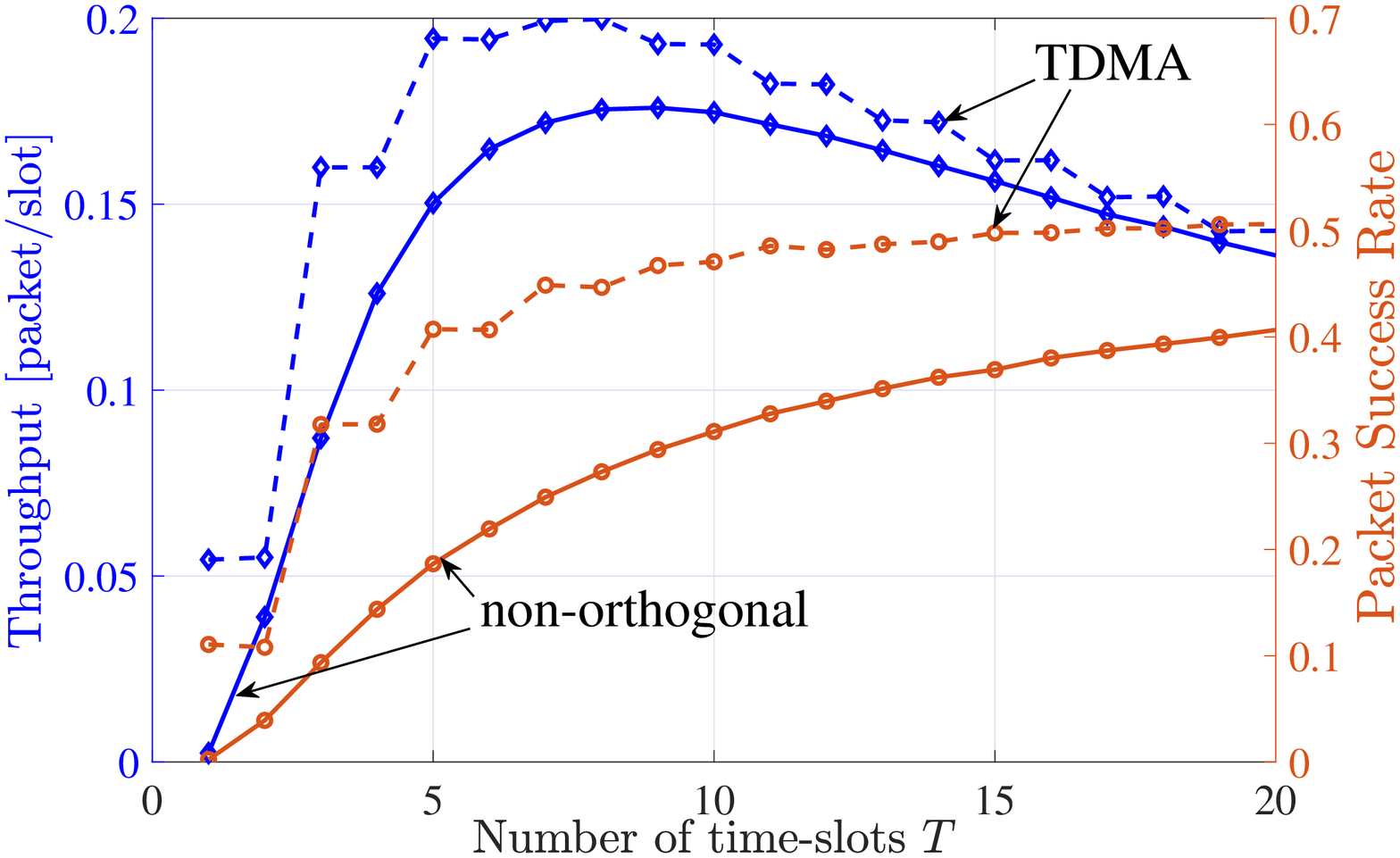}
  \caption{\ac{NCS} throughput and packet success rate}
  \label{fig:throughput_reliability_Noncritical _messages}
\end{subfigure}
\caption{\ac{CS} and \ac{NCS} throughputs and packet success rate levels as function of the number of time-slots $T$ for non-orthogonal resource allocation (solid lines) and inter-service TDMA (dashed lines) ($G=15\ [\mathrm{packet/frame}]$, $\epsilon_1 = \epsilon_2 = 0.5, L=3$ APs, $\alpha=0.5$ and $\gamma_c = 0.5$).}
\label{fig:throughput_reliability}
\end{figure*}
Finally, we consider the interplay between the throughputs and packet success rate levels for both non-orthogonal resource allocation and TDMA as function of the number of time-slots $T$. These are plotted in Fig.~\ref{fig:throughput_reliability} for $G=15\ [\mathrm{packet/frame}]$, $\epsilon_1 = \epsilon_2 = 0.5, L=3$ APs, $\alpha=0.5$ and $\gamma_c = 0.5$. For both services, following the discussion around Fig.~\ref{fig:single_service} we observe that the packet success rate level under both allocation schemes increases as function of $T$. This is because larger value of $T$ decrease chances of packet collisions. However, this is not the case for the throughput, since large values of $T$ may cause some time-slots to be left unused, which penalizes the throughput. For the \ac{CS} in Fig.~\ref{fig:throughput_reliability_critical_messages}, it is seen that non-orthogonal resource allocation outperforms TDMA in both throughput and packet success rate level due to the larger number of available resources. In contrast, Fig.~\ref{fig:throughput_reliability_Noncritical _messages} shows that TDMA provides better \ac{NCS} throughput and packet success rate level than non-orthogonal resource allocation. The main reason for this is that the lower number of resources in TDMA is compensated by the absence of inter-service interference for \ac{NCS} messages.
\section{Throughput and packet success rate Analysis Under Fading Channels}
\label{sec:throughput_reliability_fading}
The binary erasure channel model discussed in the previous sections offers a tractable set-up that facilitates the analysis of the throughput and packet success rate, enabling the derivation of closed-form expressions in various cases of interest. It is also of practical interest as a simplified model for mmwave channels \cite{mmwave_erasurechannels} and satellite communications scenarios represented in Fig.~\ref{fig:system_model}. In this section, we briefly study a more common scenario that accounts for fading channels in both radio and backhaul channels. This typically represents terrestrial scenarios as shown in Fig.~\ref{fig:system_model_earth}. More complex models that include both fading and erasures \cite{fading_and_erasure} can also be analyzed following the same steps presented below (see Section \ref{sec:conclusions} for some details). We first detail the channel and signal models, and then we derive the throughput and packet success rate metrics.
\subsection{Channel and Signal Models}
\label{sec:fading_model}
At any time-slot $t$, the channels between each user $m$ and AP $l$ and between each AP $l$ and the BS are assumed to follow the standard Rayleigh fading model, and are denoted as $h_{l,m}(t) \sim \mathcal{CN}(0,\alpha^{2})$ and $g_{l}(t) \sim \mathcal{CN}(0,\beta^{2})$, respectively. Ensuring consistency with the erasure model, we assume that all channels are independent and that the average channel gains $\alpha^2$ and $\beta^2$ are fixed. Furthermore, as detailed below, we assume that each AP and the BS decode at most one packet in each slot. Finally, we denote the transmission rates of \ac{CS} and \ac{NCS} messages as $r_c$ and $r_{\bar{c}}\ \mathrm{bit/s/Hz}$, respectively. Assuming that both access and backhaul channels are allocated the same amount of radio resources, the transmission rates are the same for both channels.\par
Given the numbers $N_c(t) = n_c$ and $N_{\bar{c}} (t)= n_{\bar{c}}$ of \ac{CS} and \ac{NCS} messages in the given time-slot $t$, the signal received at the $l$-th AP as time-slot $t$ can be written as
\begin{equation}
    y_{l}(t) = \sum_{m=1}^{n_c} h_{lm}(t)x_m (t) + \sum_{m^{\prime}=n_c+1}^{n_c + n_{\bar{c}}} h_{lm^{\prime}}(t)x_{m^{\prime}} (t) + n_l (t),
\end{equation}
where $n_l (t) \sim \mathcal{CN}(0,1)$ denotes complex white Gaussian noise at the $l$-th AP. The powers of \ac{CS} and \ac{NCS} devices are respectively denoted as
\begin{equation}
    \mathbb{E}[|x_{m}(t)|^{2} ]= P_c \label{eq:power_constraint}\ 
\mathrm{and}\ \mathbb{E}[|x_{m^{\prime}}(t)|^{2}] = P_{\bar{c}},
\end{equation}
where we take $P_c \geq P_{\bar{c}}$ to capture the generally larger transmission power of \ac{CS} transmissions. The Signal-to-Interference-plus-Noise Ratio (SINR) of message $m$ at AP $l$ is given as
\begin{equation}
    \mathrm{SINR}_{l,m}^{AP} = \frac{|h_{lm}(t)|^2 P_m}{1+ \sum_{\substack{m^\prime = 1 \\ m^{\prime} \neq m}}^{n_c + n_{\bar{c}}} |h_{lm^{\prime}}(t)|^2 P_{m^\prime}} \label{eq:SINR_l},
\end{equation}
where $P_m = P_c$ for a \ac{CS} message $m \in \{1,\ldots,n_c \}$ and $P_m = P_{\bar{c}}$ for a \ac{NCS} message $m \in \{n_c + 1 , \ldots , n_c + n_{\bar{c}} \}$. Let $m_{l}^{\star}$ denote the message with the largest SINR at the $l$-th AP, i.e., 
\begin{equation}
        m_{l}^{\star} = \argmax_{m \in \{1,\ldots , n_c + n_{\bar{c}} \}} \mathrm{SINR}_{l,m}^{AP}. \label{eq:m_lstar}
\end{equation}
The $l$-th AP only attempts to decode message $m_{l}^{\star}$. Decoding is correct if the standard Shannon capacity condition $\mathrm{SINR}_{l,m^{\star}_{l}} \geq 2^{r_{m}} -1$ is satisfied, where $r_m = r_c$ if $m_{l}^{\star} \in \{1,\ldots,n_c \}$ and $r_m = r_{\bar{c}}$ if $m_{l}^{\star} \in \{n_c,\ldots,n_c + n_{\bar{c}} \}$. \par
Each $l$-th AP transmits the decoded message $m_l^{\star}$, if any, to the BS over the wireless backhaul channel with transmission power $P_{m^{\star}_l}^{AP} = P_c^{AP}$ if $m^{\star}_{l} \in \{1,\ldots,n_c \}$ and $P_{m^{\star}_l}^{AP} = P_{\bar{c}}^{AP}$ if $m_l^{\star} \in \{n_c+1, \ldots, n_c + n_{\bar{c}}\}$. Consequently, the signal $y_{BS}(t+1)$ received at the BS in time-slot $t+1$ can be written as the sum of messages sent by all APs as
\begin{equation}
    y_{BS}(t+1) = \sum_{l=1}^{L} g_l (t+1) x_{m_l^{\star}} (t) + n_{BS}(t+1),
\end{equation}
where $n_{BS}(t) \sim \mathcal{CN}(0,1)$ denotes the white Gaussian noise at the BS.
Let $\mathcal{L}_{m} = \{l : m_l^{\star} = m \}$ denote the set of indices of APs that decoded a message $m \in \mathcal{M}^{\star}$, where $\mathcal{M}^{\star}= \{ m : \exists\ l=1,\ldots,L\ \mathrm{s.t.}\ m=m^{\star}_l\}$ denotes the set of messages decoded by at least one AP in time-slot $t$. The SINR of a message $m\in \mathcal{M}^{\star}$ received at the BS can be written as 
\begin{equation}
        \mathrm{SINR}_{m}^{BS} = \frac{|\sum_{l \in \mathcal{L}_{m}} g_{l}(t+1)|^2P^{AP}_{m} }{1 + \sum_{m^{\prime} \in \mathcal{M}^{\star} \setminus \{ m\}} |\sum_{l \in \mathcal{L}_{m^\prime}} g_l(t+1)|^2 P^{AP}_{m^\prime}}. \label{eq:SINR_BS}
\end{equation}
In a manner similar to APs, the BS attempts decoding only of the message $m^{\star}_{BS}$ with the highest SINR, namely
\begin{equation}
    m_{BS}^{\star} = \argmax_{m \in \mathcal{M}^{\star}} \mathrm{SINR}_{m}^{BS}.
\end{equation}
Message $m^{\star}_{BS}$ is decoded correctly if the standard Shannon capacity condition $\mathrm{SINR}_{m^{\star}_{BS}} \geq 2^{r_{m^{\star}_{BS}}} -1$ is satisfied, where $r_{m^\star_{BS}} = r_c$ if $m^{\star}_{BS} \in \{1,\ldots,n_c \}$ and $r_{m^{\star}_{BS}} = r_{\bar{c}}$ if $m^{\star}_{BS} \in \{n_c + 1,\ldots,n_c + n_{\bar{c}} \}$.
\subsection{Performance Analysis}
The analysis follows the same steps as in Section \ref{sec:heterogeneous_superposition}, as long as one properly redefines the probabilities $p_c$ and $p_{\bar{c}}$ of decoding correctly a \ac{CS} or a \ac{NCS} message at any given AP, as well as the probabilities $q_c$ and $q_{\bar{c}}$ of decoding correctly a \ac{CS} or \ac{NCS} message at the BS. According to the discussion in Section \ref{sec:fading_model}, the former probabilities can be respectively written as
\begin{subequations}
\begin{alignat}{1}
    & p_c = \mathrm{Pr}[m_l^{\star} \in \{1,\ldots,n_c \}\ \mathrm{and}\ \mathrm{SINR}^{AP}_{l,m^{\star}_l} > 2^{r_c}-1]\\
    & \mathrm{and}\ p_{\bar{c}} = \mathrm{Pr}[m_l^{\star} \in \{n_c + 1,\ldots,n_c + n_{\bar{c}} \}\ \\& \ \ \ \ \ \ \ \ \ \ \ \ \ \ \ \ \ \ \ \ \ \ \ \ \ \ \ \ \ \ \  \mathrm{and}\  \mathrm{SINR}^{AP}_{l,m^{\star}_l} > 2^{r_{\bar{c}}}-1],\nonumber
    \end{alignat}\label{eq:p_fading}
\end{subequations}
where $m_l^{\star}$ is defined in \eqref{eq:m_lstar}, while the latter probabilities can be redefined as 
\begin{subequations}
\begin{alignat}{1}
    & q_c = \mathrm{Pr}[m_{BS}^{\star} \in \{1,\ldots,n_c \}\ \mathrm{and}\ \mathrm{SINR}^{BS}_{m^{\star}_{BS}} > 2^{r_c}-1]\\
    & \mathrm{and}\ q_{\bar{c}} = \mathrm{Pr}[m_{BS}^{\star} \in \{n_c + 1,\ldots,n_c + n_{\bar{c}} \}\ \\& 
    \ \ \ \ \ \ \ \ \ \ \ \ \ \ \ \ \ \ \ \ \ \ \ \mathrm{and}\ \mathrm{SINR}^{BS}_{m^{\star}_{BS}} > 2^{r_{\bar{c}}}-1].
    \end{alignat} \label{eq:q_fading}
\end{subequations}
While closed-form expressions for \eqref{eq:p_fading} and \eqref{eq:q_fading} appear prohibitive to derive (see, e.g.,  \cite{outage_computation}), these probabilities can be easily estimated via Monte Carlo Simulations. Having computed probabilities \eqref{eq:p_fading}-\eqref{eq:q_fading}, the throughputs of \ac{CS} and \ac{NCS} messages can be respectively obtained using \eqref{eq:R_c_sup} and \eqref{eq:R_barc_sup}. The packet success rate can be computed by redefining $q_{c}$ and $q_{\bar{c}}$ to take into account a single message sent by a single user (for instance, the first one) as follows:
\begin{subequations}
\begin{alignat}{1}
    & q_c = \mathrm{Pr}[m_{BS}^{\star} = 1\ \mathrm{and}\ \mathrm{SINR}^{BS}_{m^{\star}_{BS}} > 2^{r_c}-1| N_c(t) \geq 1]\\
    & \mathrm{and}\ q_{\bar{c}} = \mathrm{Pr}[m_{BS}^{\star}=1\ \mathrm{and}\ \mathrm{SINR}^{BS}_{m^{\star}_{BS}} > 2^{r_{\bar{c}}}-1|N_{\bar{c}}(t) \geq 1].
    \end{alignat} \label{eq:reliability_fading}
\end{subequations}
\subsection{Examples}
\begin{figure}[h]
	\centering
	\includegraphics[height= 6.3 cm, width= 8.5 cm]{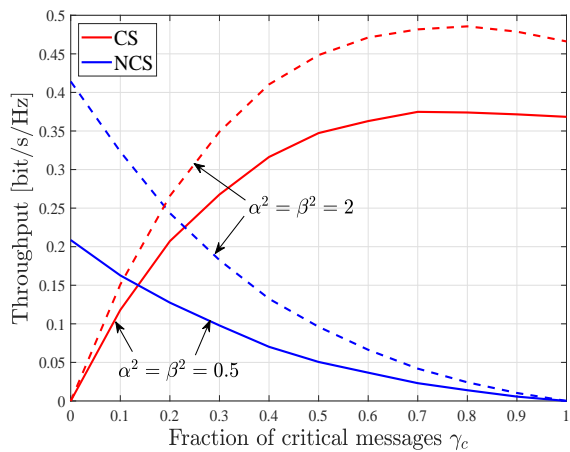}
	\caption{\ac{CS} and \ac{NCS} throughput as function of the fraction of \ac{CS} messages $\gamma_c$ under the fading channels model and using non-orthogonal resource allocation ($G=20\ [\mathrm{packet/frame}]$, $T=4\ [\mathrm{time\textrm{-}slot/frame}]$, $P_c = P^{AP}_c=10$, $P_{\bar{c}} = P_{\bar{c}}^{AP} = 4$ and $L = 3$ APs).}
	\label{fig:throuput_gammac_fading}
\end{figure}
\begin{figure}[h]
	\centering
	\includegraphics[height= 6.3 cm, width= 8.5 cm]{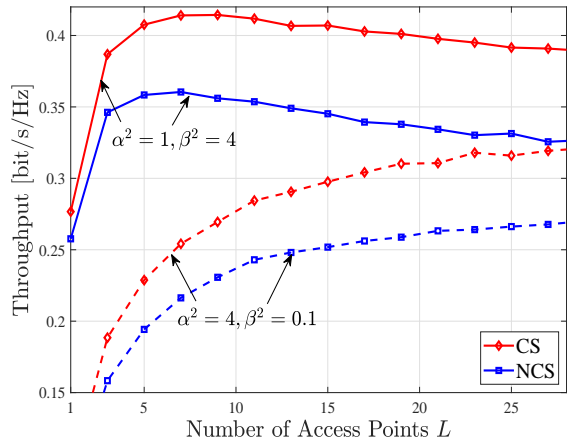}
	\caption{\ac{CS} and \ac{NCS} throughputs as function of the number of APs $L$ under the fading model and under non-orthogonal resource allocation ($G=10\ [\mathrm{packet/frame}]$, $T=4\ \mathrm{[time\textrm{-}slot/frame]}$, $\gamma_c = 0.5$, $P_c = P^{AP}_c=10$, and $P_{\bar{c}} = P_{\bar{c}}^{AP} = 9$).}
	\label{fig:rate_L_fading}
\end{figure}
We now consider the fading channels model discussed in Sec. \ref{sec:throughput_reliability_fading} with the main aim of relating the insights obtained from the analysis of erasure channels to the more common Rayleigh fading setup. We fix $P_c = P^{AP}_c = 10$ and $T = 4\ [\mathrm{time\text{-}slot/frame}]$. In an analogy to Fig.~\ref{fig:rate_load_1}, we start in Fig.~\ref{fig:throuput_gammac_fading} by investigating the throughput of \ac{CS} and \ac{NCS} messages as function of the fraction of \ac{CS} messages $\gamma_c$ for different values of the average channel powers $\alpha^2$ and $\beta^2$. In general, we observe similar trends as in Fig.~\ref{fig:rate_load_1}. Most notably, the throughput of \ac{CS} messages peaks at a value of $\gamma_c$ that strikes the best balance between the combining gains due to the transmission of a message from multiple APs and the interference created by concurrent AP transmissions of different messages. However, in contrast to the erasure model in which increasing the erasure rate can be advantageous, the throughput of both services improves as the average channel strengths $\alpha^2$ and $\beta^2$ are increased. This is because interference from concurrent transmissions has a more deleterious effect under the collision model assumed when considering erasures than under the SINR model. For the latter, reducing both channel strengths $\alpha^2$ and $\beta^2$ has the net effect of reducing the SINRs despite the decrease in interference power.\par
Finally, to compare some of the design insights from the erasure model, we plot in Fig.~\ref{fig:rate_L_fading} the throughput of both services as function of the number of APs $L$. We can see that, in a manner similar to Fig. ~\ref{fig:rate_L}, when $\alpha^2$ is high and $\beta^2$ is low, which is akin to lower $\epsilon_1$ and high $\epsilon_2$ in Fig.~\ref{fig:rate_L}, the throughput of both services increases as function of $L$. This is because low values of the channel power $\beta^2$ in the backhaul channel imply that the SINR is limited by the signal power and not by interference. Therefore, increasing the space diversity via a larger $L$ can be advantageous in this regime. Furthermore, as evinced from Fig.~\ref{fig:rate_L}, the SINR of the backhaul channel may be limited by the level of interference and hence when $\alpha^2$ is low and $\beta^2$ is high, which is akin to high $\epsilon_1$ and low $\epsilon_2$ in Fig.~\ref{fig:rate_L}, increasing the number of APs beyond a given threshold reduces the throughput.
\section{Conclusions And Extensions}
\label{sec:conclusions}
This paper studies grant-free random access for coexisting \ac{CS} and \ac{NCS} in IoT systems with shared wireless backhaul and uncoordinated APs. 
Non-orthogonal and orthogonal inter-service resource sharing schemes based on random access are considered. From the \ac{CS} perspective, it was found that non-orthogonal sharing is preferable to a standard inter-service TDMA protocol in terms of both throughput and packet success rate level. In contrast, this is not the case for the \ac{NCS}, since inter-service orthogonal resource allocation eliminates interference from the larger-power \ac{CS}. 
Furthermore, both erasure and fading channels models are considered to capture satellite and terrestrial applications respectively. Similarities were found between these models which proves the suitability of the erasure model for such type of analysis due to its mathematical tractability properties. Through extensive numerical results, the impact of both spatial and time resources is investigated, revealing trade-offs between throughput and packet success rate for both services.\par
\textcolor{black}{Our model could be directly extended to cater for multiple antennas satellites, known as multi-beam satellites, and more realistic inter-satellite interference. For the former, different beams could be used per-antenna to cover different areas, with the proposed grant-free NOMA applied separately to each area. As for the latter, it is more realistic to assume different direct and interference channel gains to account for the revolution of LEO satellites around the planet. This can be directly considered by modeling interference channels with a different channel gain than the direct channel. The analysis could be directly extended to this case at the price of a more cumbersome notation. } 
\appendix

\section{}
\label{app:tru}

\subsection{Proof of Proposition 3}
An AP successfully retrieves a non-critical packet when only one of the $\NTxNC$ non-critical packets transmitted arrives unerased and all critical transmitted packets $\NTx $ do not reach the AP due to erasures. This happens with probability 
\begin{equation}
    {p_{N_{\bar{c}}} = \underbrace{n_{\bar{c}} \,(1-\perasU) \, \perasU^{n_{\bar{c}}-1}}_{(a)}  \underbrace{\perasU^{n_c}}_{(b)} }, \label{eq:p_nbarc}
\end{equation}
where term $(a)$ is the probability that one of the $n_{\bar{c}}$ non-critical messages is successfully received at an AP and term $(b)$ is the probability all critical messages are erased. On the backhaul, a non-critical packet reaches the BS via one of the APs when the packet is not erased over the backhaul channel and no other packet (critical and non-critical) is successfully received from any of the remaining $\nRx-1$ APs. The overall probability of successful reception at the BS is thus $q_{\bar{c}} = \nRx q_{n_{\bar{c}}}(1-q_{n_c}-q_{n_{\bar{c}}})^{\nRx-1}$, with $q_{n_{\bar{c}}} = p_{n_{\bar{c}}}  \,(1-\perasD)$ and $q_{n_c}$ defined in \eqref{eq:q_nc_single_Service}. The non-critical throughput can then be written by averaging $q_{\bar{c}}$ over $n_c$ and $n_{\bar{c}}$ as
\begin{equation}
\begin{aligned}
  R_{\bar{c}} = \mathbb{E}_{N_c, N_{\bar{c}}}[q_{\bar{c}}]= \sum_{\nTxNC=0}^{\infty} \sum_{\nTx=0}^{\infty} & \frac{G_{\bar{c}}^\nTxNC e^{-G_{\bar{c}}}}{\nTxNC!} \frac{\load^\nTx e^{-\load}}{\nTx!} \\& \cdot \nRx \, \psDNC (1-\psD-\psDNC)^{\nRx-1}.
\label{eq:truSumNC}
\end{aligned}
\end{equation}
Following similar steps as the one detailed in the proof of \textit{Proposition 1} and after some tedious yet straightforward rearrangements, the non-critical throughput in \eqref{eq:truSumNC} can be written in closed form as detailed in \eqref{eq:R_barc_no_K_collision}.
\par We now derive the non-critical packet success rate. Similar to the single-service case, the probability of receiving a non-critical packet at an AP from a given user $u$ given that $u$ is active is given by 
\begin{equation}
    p_u = (1-\epsilon_1)\epsilon_1^{n^\prime_{\bar{c}}-1}\epsilon_1^{n_c}. \label{eq:p_u_non_c_K_infinite}
\end{equation}
The probability that the packet from user $u$ is received successfully at the BS is then computed as 
\begin{equation}
    q_u = L p_u (1-\epsilon_2) (1-q)^{L-1},
\end{equation}
where $q = (1-\epsilon_2)[p_{n_c} + p_{n^{\prime}_{\bar{c}}}]$ is the probability that the BS successfully receives a critical message or a non critical message from any of the remaining $L-1$ APs. The non-critical packet success rate $\Gamma_{\bar{c}}$ can be obtained by averaging $q_u$ over $n^\prime_{\bar{c}}$ and $n_c$, where $P[N^\prime_{\bar{c}}=n^\prime_{\bar{c}}| N^\prime_{\bar{c}} \geq 1] = (1-e^{-G_{\bar{c}}})^{-1} (e^{-G_{\bar{c}}}G_{\bar{c}}^{n_{\bar{c}}^{\prime}})/(n^{\prime}_{\bar{c}}!)$ is the distribution of non-critical packets given that user $u$ is active. The non-critical packet success rate $\Gamma_{\bar{c}}$ can be obtained in closed form by following similar steps in the proof of \textit{Proposition 1}.
\subsection{Proof of Proposition 4}
The probability of receiving a non-critical message at the BS can be written as 
\begin{equation}
    q_{\bar{c}} = L q_{n_{\bar{c}}} (1-q_{n_c} - q_{n_{\bar{c}}})^{L-1}, \label{eq:q_barc_finite_K}
\end{equation}
where $q_{n_c} = p_{n_c} \gamma_K(n_{\bar{c}}, \epsilon_1) (1-\epsilon_2)$ and $q_{n_{\bar{c}}} = p_{n_{\bar{c}}}(1-\epsilon_2)$ are the probabilities of receiving successfully a critical or non-critical packet respectively at the BS.
The non-critical throughput $R_{\bar{c}}$ can be then obtained by averaging $q_{\bar{c}}$ in \eqref{eq:q_barc_finite_K} over all values of $n_c$. \\
Moving to non-critical packet success rate $\Gamma_{\bar{c}}$, the probability of receiving successfully the packet of a given user $u$ at an AP is defined in the same way as in \eqref{eq:p_u_non_c_K_infinite}. The probability of receiving this packet at the BS can be written as 
\begin{equation}
    q_{\bar{c}} = L p_u (1-\epsilon_2) (1-q)^{L-1},
\end{equation}
where $q$ is the probability of receiving any critical or non-critical packet from the remaining $L-1$ APs. This can be written as $q = (1-\epsilon_2) [p_{n_c} \gamma_{K-1}(n^{\prime}_{\bar{c}}, \epsilon_1) + n^{\prime}_{\bar{c}} p_u]$, where the first part corresponds to receiving any critical message at the BS while the second part to receiving any non-critical message at the BS. The non-critical packet success rate $\Gamma_{\bar{c}}$ can be then obtained by averaging $q_{\bar{c}}$ over $N_c = n_c$ and $N_{\bar{c}} = n_{\bar{c}}$. \par
\subsection{Proof of Proposition 5}
The event that a transmitted critical packet is received at the BS passing through one of the APs occurs if the AP successfully decodes one critical packet, and the packet is not erased over the backhaul channel. Conditioned on $\NTx=\nTx$ and $\NTxNC=\nTxNC$, this event has the probability $\psDP=\psU\gamma_K(n_{\bar{c}}, \epsilon_1)\,(1-\perasD)$. The number of incoming backhaul critical packets over a slot follows the distribution $\mathrm{Bin}(\nRx,\psDP)$. Hence, the critical throughput can be written as  
\begin{equation}
R_c = \sum_{\nTx=0}^{\infty}\sum_{\nTxNC=0}^{\infty} \frac{\load^\nTx e^{-\load}}{\nTx!} \frac{\loadNC^\nTxNC e^{-\loadNC}}{\nTxNC!} \cdot \nRx \, \psDP (1-\psDP)^{\nRx-1}.
\label{eq:truDKfiniteSum}
\end{equation}
Now we split the sum over the non-critical packets transmitted $\nTxNC$ in two parts, the first considers a number of non-critical packets not exceeding $n_{\bar{c}} < K$, while the second part corresponds ${\nTxNC\geq K+1}$. In the first part $\gamma_K(n_{\bar{c}}, \epsilon_1)=1$ by definition, so $\psDP=\psD$. Consequently, \eqref{eq:truDKfiniteSum} can be written as the sum of two terms
\begin{align}
\begin{split}
R_c &= \sum_{\nTxNC=0}^{K}\frac{\loadNC^\nTxNC e^{-\loadNC}}{\nTxNC!} \sum_{\nTx=0}^{\infty}\frac{\load^\nTx e^{-\load}}{\nTx!}  \cdot \nRx \, \psD (1-\psD)^{\nRx-1}\\ &+ \sum_{\nTxNC=K+1}^{\infty}\frac{\loadNC^\nTxNC e^{-\loadNC}}{\nTxNC!} \sum_{\nTx=0}^{\infty}\frac{\load^\nTx e^{-\load}}{\nTx!}  \cdot \nRx \, \psDP (1-\psDP)^{\nRx-1}.
\end{split}
\label{eq:truDKfiniteSumSplit}
\end{align}
The first term in \eqref{eq:truDKfiniteSumSplit} is the product between the \ac{CDF} of a Poisson distribution of parameter $\loadNC$ computed in $K$ and the same expression of the throughput for the single service case found in Section \ref{sec:oneclass} in \eqref{eq:truSum}. As for the second term, following a simple yet tedious mathematical derivation it can be written as
\begin{align}
\begin{split}
&\sum_{\nTxNC=K+1}^{\infty}\frac{\loadNC^\nTxNC e^{-\loadNC}}{\nTxNC!} \sum_{\nTx=0}^{\infty}\frac{\load^\nTx e^{-\load}}{\nTx!}  \cdot \nRx \, \psDP (1-\psDP)^{\nRx-1}\\ &= \sum_{\ell=0}^{\nRx -1} (-1)^\ell \, \nRx \, {\nRx-1 \choose \ell} \left[\frac{ (1-\perasU) (1-\perasD)}{\perasU}\right]^{\ell+1}  e^{-\load} \\ &\ \ \ \ \ \ \ \ \ \ \ \ \ \ \ \ \ \ \ \ \ \ \ \ \ \cdot \,\,\ancF_{\ell+1}\left(\load \,\perasU^{\ell+1}\right) \cdot \,\xi(K,\loadNC,\ell),
\end{split}
\label{eq:finiteSum}
\end{align}
where 
\begin{equation}
    \xi(K,\loadNC,\ell)\!=\!\!\sum_{\nTxNC=K+1}^{\infty} \!\!\!\frac{\loadNC^\nTxNC \, e^{-\loadNC}}{\nTxNC!}\!\left[ \sum_{i=0}^{K} {\nTxNC \choose i} \left(1-\perasU \right)^i \perasU^{\nTxNC - i}\right]^{\ell+1}.
\end{equation} Finally, putting together \eqref{eq:truDKfiniteSumSplit} and \eqref{eq:finiteSum} the lemma can be concluded.\\
Moving to the packet success rate, the probability of receiving a given critical packet from a user $u$ at an AP is 
\begin{equation}
    p_u = (1-\epsilon_1)\epsilon_1^{n^\prime_c-1} \gamma_K(n_{\bar{c}}, \epsilon_1). \label{eq:p_u_small_L}
\end{equation}
The probability of receiving this critical packet at the BS is given by
\begin{equation}
    q_c = L p_u (1-\epsilon_2) (1-q_{n^{\prime}_c})^{L-1} \label{eq:q_c_1}
\end{equation}
where $q_{n^{\prime}_c} = n^{\prime}_c (1-\epsilon_1) \epsilon_1^{n^{\prime}_{c}-1} \gamma_K(n_{\bar{c}}, \epsilon_1) (1-\epsilon_2)$ is the probability of receiving any critical packet at the BS. Finally the critical packet success rate can be obtained by averaging $q_c$ over $n^{\prime}_c$ and $n_{\bar{c}}$.

\subsection{Asymptotic Throughput and Packet Success Rate for Single Service}
We now state two theorems regarding the asymptotic behaviour of the throughput and packet success rate for large number of APs.\par
\textit{Theorem 1: } \textit{The critical throughput tends to zero for large number of APs, i.e., $ \underset{L \to +\infty}{\lim} R_c =  0$.}\\
\textit{Proof: }
Let us first start by defining $R_c(m) = \sum_{n_c=1}^{m} \frac{G_{c}^{n_c} e^{-G_{c}}}{n_c!} \cdot L q_{n_c} (1-q_{n_c})^{L-1}$, which represents the summation in \eqref{eq:rate_single_service} up to $m$ terms. Note that the summation starts at $n_c = 1$ as the throughput is null otherwise. Consequently, we have 
\begin{align}
    \underset{L \to \infty}{\lim} R_c &= \underset{L \to \infty}{\lim} \underset{m \to \infty}{\lim} R_c (m) \nonumber\\
    & \underset{(a)}{=}  \underset{m \to \infty}{\lim} \underset{L \to \infty}{\lim}R_c (m) \nonumber\\
    & = \underset{m \to \infty}{\lim} \sum_{n_c=1}^{m} \frac{G_{c}^{n_c} e^{-G_{c}}}{n_c!} \cdot \underset{L \to \infty}{\lim} \big[ L q_{n_c} (1-q_{n_c})^{L-1} \big] \nonumber \\
    & \underset{(b)}{=} 0, \nonumber
\end{align}
where $(a)$ follows from Moore-Osgood theorem \cite[Theorem 1]{Moore_osgood} for interchanging limits using the fact that $R_c (m)$ converges for each value of $L$ and $(b)$ is due to the fact that the second limit is null because $(1-q_{n_c}) < 1$. \qed
\\
\textit{Theorem 2: } \textit{The critical packet success rate tends to zero for large number of APs, i.e., $ \underset{L \to +\infty}{\lim} \Gamma_c =  0$.}\\
\textit{Proof: }The proof follows similar steps as the proof for Theorem 1 detailed above.\qed \\

\bibliographystyle{IEEEtran}
\bibliography{Biblio} 
\end{document}